\documentclass[letters,fleqn,usenatbib]{mnras}

\usepackage{newtxtext,newtxmath}

\usepackage[flushleft]{threeparttable} 
\usepackage{soul}

\usepackage{amssymb}
\usepackage{lscape}
\usepackage{comment}
\usepackage{ulem}
\usepackage{makecell}
\usepackage{bm}
\usepackage{pdflscape} 
 
\usepackage{hyperref}
\hypersetup{colorlinks=true,linkcolor=blue,citecolor=blue,filecolor=blue,urlcolor=blue}

\usepackage{multirow}

\usepackage[T1]{fontenc}

\DeclareRobustCommand{\VAN}[3]{#2}
\let\VANthebibliography\thebibliography
\def\thebibliography{\DeclareRobustCommand{\VAN}[3]{##3}\VANthebibliography}

\newcommand{\targetone}{EP241103a}
\newcommand{\targettwo}{EP260409a}
\newcommand{\tzerotone}{T$_{0}^{1103\mathrm{a}}$}
\newcommand{\tzerottwo}{T$_{0}^{0409\mathrm{a}}$}

\usepackage{graphicx}	
\usepackage{amsmath}	

\title[Dark FXTs]{Gas absorption of soft X-rays strongly impacts the redshift distribution of dark Fast X-ray Transients}

\author[Sánchez-Sierras et al.]{Javier Sánchez-Sierras,$^{1}$\thanks{E-mail: javisansie@gmail.com}
Peter G. Jonker,$^{1}$
Jonathan Quirola-V\'asquez,$^{1}$
Agnes P. C. van Hoof,$^{1}$
\newauthor Andrew J. Levan,$^{1,2}$
Maria E. Ravasio,$^{3,4,1}$
Joyce N. D. van Dalen,$^{1}$
Franz E. Bauer,$^{5}$
Jia-Ying Cao,$^{6}$
\newauthor Francesco Carotenuto,$^{7}$
Jennifer A. Chacón,$^{8}$
Ashley A. Chrimes,$^{9,1}$
Laura Cotter,$^{10}$
Gregory Corcoran,$^{10}$
\newauthor Rob A. J. Eyles-Ferris,$^{11}$
Guoli Huang,$^{6,12}$
Shuai-Qing Jiang,$^{13,14}$
Zexi Li,$^{6,12}$
Yifang Liang,$^{15,16}$
\newauthor He-Yang Liu,$^{13}$
Daniele B. Malesani,$^{17}$
Antonio Martin-Carrillo,$^{10}$
Daniel Mata S\'anchez,$^{18,19}$
\newauthor Nikhil Sarin,$^{20,21}$
Manuel A. P. Torres,$^{18}$
Qinyu Wu,$^{13,14}$
Tong Zhao,$^{22,14}$
and Wen-Da Zhang$^{13}$
\\
$^{1}$Department of Astrophysics/IMAPP, Radboud University, Heyendaalseweg 135, NL-6525 AJ Nijmegen, The Netherlands\\
$^{2}$Department of Physics, University of Warwick, Coventry, CV4 7AL, UK\\
$^{3}$Institute of Space Sciences (ICE), CSIC, Campus UAB, Carrer de Can Magrans s/n, Barcelona, E-08193, Spain\\
$^{4}$Institut d’Estudis Espacials de Catalunya (IEEC), Edifici RDIT, Campus UPC, Castelldefels (Barcelona), E-08860, Spain\\
$^{5}$Instituto de Alta Investigaci\'on, Universidad de Tarapac\'{a}, Casilla 7D, Arica, Chile \\
$^{6}$Key Laboratory of Particle Astrophysics, Institute of High Energy Physics, Chinese Academy of Sciences, Beijing 100049, China\\
$^{7}$INAF-Osservatorio Astronomico di Roma, Via Frascati 33, I-00078, Monte Porzio Catone, Italy\\
$^{8}$Instituto de Astrofísica, Facultad de Física, Pontificia Universidad Católica de Chile, Campus San Joaquín, Av. Vicuña Mackenna 4860, Macul Santiago, 7820436, Chile\\
$^{9}$European Space Agency (ESA), European Space Research and Technology Centre (ESTEC), Keplerlaan 1, 2201 AZ Noordwijk, the Netherlands\\
$^{10}$School of Physics and Centre for Space Research, University College Dublin, Belfield, Dublin 4, Ireland\\
$^{11}$School of Physics and Astronomy, University of Leicester, University Road, Leicester, LE1 7RH, UK\\
$^{12}$University of Chinese Academy of Sciences, Chinese Academy of Sciences, Beijing 100049, China\\
$^{13}$National Astronomical Observatories, Chinese Academy of Sciences, Beijing 100101, China\\
$^{14}$School of Astronomy and Space Sciences, University of Chinese Academy of Sciences, Datun Road A20, Beijing 100049, China\\
$^{15}$Purple Mountain Observatory, Chinese Academy of Sciences, Nanjing 210008, China\\
$^{16}$School of Astronomy and Space Sciences, University of Science and Technology of China, Hefei 230026, China\\
$^{17}$Niels Bohr Institute, University of Copenhagen, Jagtvej 155, 2200, Copenhagen N, Denmark\\
$^{18}$Instituto de Astrof\'isica de Canarias, E-38205 La Laguna, Tenerife, Spain\\
$^{19}$Departamento de Astrof\'isica, Univ. de La Laguna, E-38206 La Laguna, Tenerife, Spain\\
$^{20}$Kavli Institute for Cosmology Cambridge, Madingley Road, Cambridge CB3 0HA, UK\\
$^{21}$Institute of Astronomy, University of Cambridge, Madingley Road, Cambridge CB3 0HA, UK\\
$^{22}$Key Laboratory for Computational Astrophysics, National Astronomical Observatories, Chinese Academy of Sciences, Datun Road A20, Beijing 100012, China\\
}

\date{Accepted XXX. Received YYY; in original form ZZZ}

\pubyear{\the\year{}}

\begin{document}
\newcommand{\he}[1] {He\,{\sc #1}}
\newcommand{\hel}[2] {He\,{\sc #1}~{$\lambda$#2}}
\newcommand{\ha}{H$\alpha$}
\newcommand{\hb}{H$\beta$}
\newcommand{\hg}{H$\gamma$}
\newcommand{\hd}{H$\delta$}
\newcommand{\hep}{H$\epsilon$}
\newcommand{\hei}{He~{\sc i}}
\newcommand{\heii}{He~{\sc ii}}

%
%
\def\kms{\mbox{${\rm km}\:{\rm s}^{-1}\:$}}
\def\ms{${\rm m}\:{\rm s}^{-1}\:$}

\def\lesssim{\mathrel{\hbox{\rlap{\hbox{\lower4pt\hbox{$\sim$}}}\hbox{$<$}}}}
\let\la=\lesssim
\def\gtrsim{\mathrel{\hbox{\rlap{\hbox{\lower4pt\hbox{$\sim$}}}\hbox{$>$}}}}
\let\ga=\gtrsim
\def\fdg{\hbox{$.\!\!^\circ$}}
\newcommand{\fdeg}{\hbox{$.\!\!^\circ$}}
\def\farcs{\hbox{$.\!\!^{\prime\prime}$}}
\def\arcsec{\hbox{$^{\prime\prime}$}}
\def\kmm{${\rm km}^{-1}$}
\def\d{{\rm d}}
\def\degree{\mbox{$^{\circ}$}}
\def\sol{$~\mathrm{M}_\odot$}
\def\lx{$L_\mathrm{X}$}
\def\kem{K_\mathrm{em}}
\def\ergs{erg s$^{-1}$}
\def\ledd{$L_\mathrm{Edd}$}
\def\vt{$v_\mathrm{t}$}

\def\sloanu{$u^\prime$}
\def\sloang{$g^\prime$}
\def\sloanr{$r^\prime$}
\def\sloani{$i^\prime$}
\def\sloanz{$z^\prime$}

\def\approxlt{\ifmmode \rlap{$<$}{}_{{}_{{}_{\textstyle\sim}}} \else%
$\rlap{$<$}{}_{{}_{{}_{\textstyle\sim}}}$\fi}
\def\approxgt{\ifmmode \rlap{$>$}{}_{{}_{{}_{\textstyle\sim}}} \else%
$\rlap{$>$}{}_{{}_{{}_{\textstyle\sim}}}$\fi}
\def\fluxnospace{erg cm$^{-2}$ s$^{-1}$}
\def\lum{erg s$^{-1}$~}
\def\lumnospace{erg s$^{-1}$}
\def\msun{M_\odot}
\def\arcmin{\hbox{$^\prime$}}
\def\arcsec{\hbox{$^{\prime\prime}$}}

\label{firstpage}
\pagerange{\pageref{firstpage}--\pageref{lastpage}}
\maketitle

\begin{abstract}
The progenitors of many Fast X-ray Transients (FXTs) and those of $\gamma$-ray bursts (GRBs) are strongly linked. Given that "dark" GRBs are typically found to be events suffering from enhanced extinction in the host galaxy, we investigate the nature of "dark" FXTs. However, unlike $\gamma$-rays, soft X-rays are strongly affected by absorption, implying that dark FXTs discovered by Einstein Probe's Wide-field X-ray Telescope (EP-WXT) have to be emitted at energies $>$2--3 keV restframe hence at a redshift $z\gtrsim2$. To illustrate this we present two dark FXTs, \targetone{} and \targettwo{}, discovered by EP-WXT. For \targetone{}, the high extinction precluded the detection of the optical counterpart, while for a probable source redshift of $\approx2.5$, the rest-frame X-ray photons are not severely affected. For \targettwo{} no counterpart is detected in the optical, but we do detect a near-infrared counterpart. A plausible scenario for \targettwo{} is that it lies at a redshift $z \gtrsim 4$. Additionally, we conduct simulations of FXTs observed with EP-WXT, showing that the detected counts for absorbed events ($N_{\mathrm{H}}\gtrsim10^{22}$~cm$^{-2}$) strongly decrease at low redshifts, hindering their detection. These results support the theoretical prediction that dark FXTs have an intermediate to high redshift and a different selection function from those of dark GRBs.
\end{abstract}





\section{Introduction}

Fast X-ray transients (FXTs) are extragalactic X-ray bursts lasting from minutes to hours (e.g. \citealt{Soderberg2008,Jonker2013,Bauer2017}). FXTs were initially discovered by mining X-ray archival data from different observatories (e.g. \citealt{Quirola-Vasquez2022,Quirola-Vasquez2023a,Zand2025,Brightman2026}). The latency of these discoveries impeded multiwavelength detections of these events. However, this changed with the launch of the Einstein Probe satellite (EP; \citealt{Yuan2022,Yuan2025}) in January 2024. The rapid alerts with latencies of a few minutes to hours for newly detected FXTs allowed multiwavelength detections (e.g. \citealt{Aryan2025}), which enabled to study their origins more extensively. FXTs have been frequently linked to $\gamma$-ray bursts (GRBs), since some FXTs have $\gamma$-ray counterparts (e.g.~\citealt{Zhang2024b,Jiang2025}) and some proposed formation channels are common to both (e.g. \citealt{Guo2025}). Several FXTs have been associated to Type Ic broad-line supernovae \citep[e.g.][]{vanDalen2025,Eyles-Ferris2025,Srinivasaragavan2025}, which are also commonly associated with long GRBs. However, there have also been links to short GRBs (e.g. EP250704a/GRB~250704B; \citealt{Swain2026,Fraija2026}).

As a result of the growing number of FXT discoveries, a subpopulation lacking detected optical counterparts, despite of the early ($\lesssim1$~day after the discovery) and deep follow-up observations, has emerged. These so-called dark FXTs (e.g. \citealt{Aryan2025}) resemble the population of dark GRBs, for which optical emission also remains undetected in early GRB follow-up. The dark GRB fraction of the whole GRB population has been estimated to be $\sim 25 - 40\%$ \citep{Perley2016}. \cite{Jakobsson2004} defined a criterion to evaluate the darkness of a GRB using the optical-to-X-ray spectral index ($\beta_{\mathrm{OX}}$; $F_{\nu} \propto \nu^{-\beta_{\mathrm{OX}}}$) of the afterglow emission, which must satisfy $\beta_{\mathrm{OX}} < 0.5$ to be considered dark. This was refined by \cite{vanderHorst2009}, who also considered the X-ray emission through the X-ray spectral index ($\beta_{\mathrm{X}}$) in the definition of optically dark GRBs, requiring to fulfill the condition $\beta_{\mathrm{OX}} < \beta_{\mathrm{X}} - 0.5$. Diverse scenarios have been proposed to explain the optical darkness of dark GRBs: (i) the optical emission is absorbed by host galaxy material along the line of sight, (ii) the source lies at very high redshift (i.e. $z \gtrsim 5$) so that Ly$\alpha$ absorption caused by the intergalactic medium strongly suppresses the observer-frame optical flux, and (iii) the source is intrinsically optically faint \citep{Greiner2011}.

Here, we investigate the differences that exist between the $\gamma$-ray and X-ray sources and what that implies for the sample of dark FXTs compared with that of dark GRBs. We illustrate this through two example dark FXTs and their properties. The X-ray transient \targetone{} was detected by the Wide-field X-ray Telescope (EP-WXT) instrument onboard the EP satellite on 2024-11-03 01:23:37 (\tzerotone{} hereafter), and it lasted for $\sim$60~s \citep{Zhao2024}. No optical counterpart of \targetone{} was identified in two subsequent early observations in the $r^{\prime}$ band at \tzerotone{}$ + 1.9$~h and \tzerotone{}$ + 2.3$~h, with 3$\sigma$ upper limiting AB magnitudes of $r^{\prime}$>20.9 and $r^{\prime}$>21.5, respectively \citep{gcnIzzo2024,gcnLi2024}. 

\targettwo{} was detected by EP-WXT on 2026-04-09 00:16:43 (\tzerottwo{} hereafter; \citealt{gcnLi2026b}). No optical or near-infrared counterpart was detected in observations of \targettwo{} performed at \tzerottwo{}$ + 2.02$~h and \tzerottwo{}$ + 3.35$~h with 3$\sigma$ AB upper limits of $VT_{\mathrm{B}}$>23.7, $VT_{\mathrm{R}}$>23.6 \citep{gcnLi2026a}, and $J$>19.5, $H_{\mathrm{s}}$>19.8 \citep{gcnKarambelkar2026}. In this work, we present X-ray and deep optical and near-infrared observations of \targetone{} and \targettwo{}. Throughout this work, we assume a flat $\Lambda$CDM cosmology with H$_{0} = 67.7$~km~s$^{-1}$~Mpc$^{-1}$ and $\Omega_{\Lambda} = 0.68$ \citep{Aghanim2020}. All magnitudes are reported in the AB system, upper limits are at the 3$\sigma$ level, error bars are at the 90\% confidence level, and dates are given in UTC.
\section{Observations and data reduction}
\label{secObservationsReduction}

\subsection{X-rays}

The EP Follow-up X-ray Telescope (EP-FXT) automatically slewed to observe the sky region of \targetone{}, starting approximately five minutes after \tzerotone{}, for a total exposure time of 2.3~ks. The source was detected at $\mathrm{RA_{J2000}} = 27.7578^{\circ}$, $\mathrm{Dec_{J2000}} = 18.9566^{\circ}$, with an uncertainty radius of 10~arcsec \citep{Zhao2024b}. Two additional EP-FXT observations were obtained. The first started at \tzerotone{}$ + 13.56$~h with an exposure time of 3.1~ks and the second started at \tzerotone{}$ + 8.58$~d with an exposure time of 2.13~ks. For \targettwo{}, two EP-FXT observations were performed, starting at \tzerottwo{}$ + \sim$5~m and \tzerottwo{}$ + 20.71$~h, with exposure times of 2.02~ks and 3.54~ks, respectively. The coordinates of \targettwo{} derived from the first observation are $\mathrm{RA_{J2000}} = 172.9096^{\circ}$, $\mathrm{Dec_{J2000}} = -9.3457^{\circ}$, with an uncertainty radius of 10~arcsec \citep{gcnLi2026b}. 

The EP-WXT data were processed with \texttt{wxtpipeline}. The X-ray light curves and spectra were extracted following standard procedures of the \texttt{XSELECT} tool \citep{Heasarc2014}. We extracted X-ray photons from a circular region with a radius of 9\arcmin\,centred on the source position, while background photons were extracted from a source-free annular region centred on the source with inner and outer radii of 11\arcmin\ and 20\arcmin, respectively. For the EP-FXT data, we reduced the observations using the FXT data analysis pipeline (\texttt{fxtsoftware} v1.20\footnote{ http://epfxt.ihep.ac.cn/analysis}), following the EP-FXT User’s Guide (v1.20). The X-ray spectrum and response files were obtained using the \texttt{XSELECT} tool, considering a circular region centred on the target with a radius of 40\arcsec. A circular region with a radius of 90\arcsec\,near the source was selected to extract the background. All X-ray spectra have at least 1~count per spectral bin, and were analysed using \texttt{XSPEC} v12.14 \citep{Arnaud1996} employing Cash statistics (\citealt{Cash1979}; W-stat because the spectra are background subtracted). 

The Neil Gehrels Swift Observatory X-ray Telescope (\textit{Swift}-XRT; \citealt{Gehrels2004,Burrows2005}) observed the field of \targetone{} at \tzerotone{} + 8.67~h \citep{Jonker2025a}. The observation lasted 984~s, and an X-ray source consistent with the EP-FXT position of \targetone{} was detected, refining the coordinates to $\mathrm{RA_{J2000}} = 27.75834^{\circ}$, $\mathrm{Dec_{J2000}} = 18.95698^{\circ}$, with an error radius of 4.6 arcsec. The data were obtained and processed with the online Build \textit{Swift}-XRT products tool \citep{Goad2007,Evans2009}.

\subsection{Optical and near-infrared}

We obtained photometric observations of the field of \targetone{} with the Gemini Multi-Object Spectrograph (GMOS; \citealt{Hook2004}; GS-2024B-Q-131, PI Bauer), installed at the 8.1-m Gemini South Telescope (GS), Cerro Pachón, Chile. The first epoch started at \tzerotone{}$ + 4.94$~h and consisted of $4\times120$~s $r^\prime$-band exposures. The second epoch started at \tzerotone{}$ + 26.10$~h and comprised both $r^\prime$- and $z^\prime$-band exposures with $5\times240$~s in $r^\prime$ and $9\times120$~s in $z^\prime$. A summary of the optical and near-infrared observations is presented in Table~\ref{tableObsDetails}. We reduced the data with the DRAGONS pipeline v4.1.0 \citep{Labrie2023,Simpson2023} under standard settings.

\begin{figure*}
    \centering
    \includegraphics[width=5truecm]{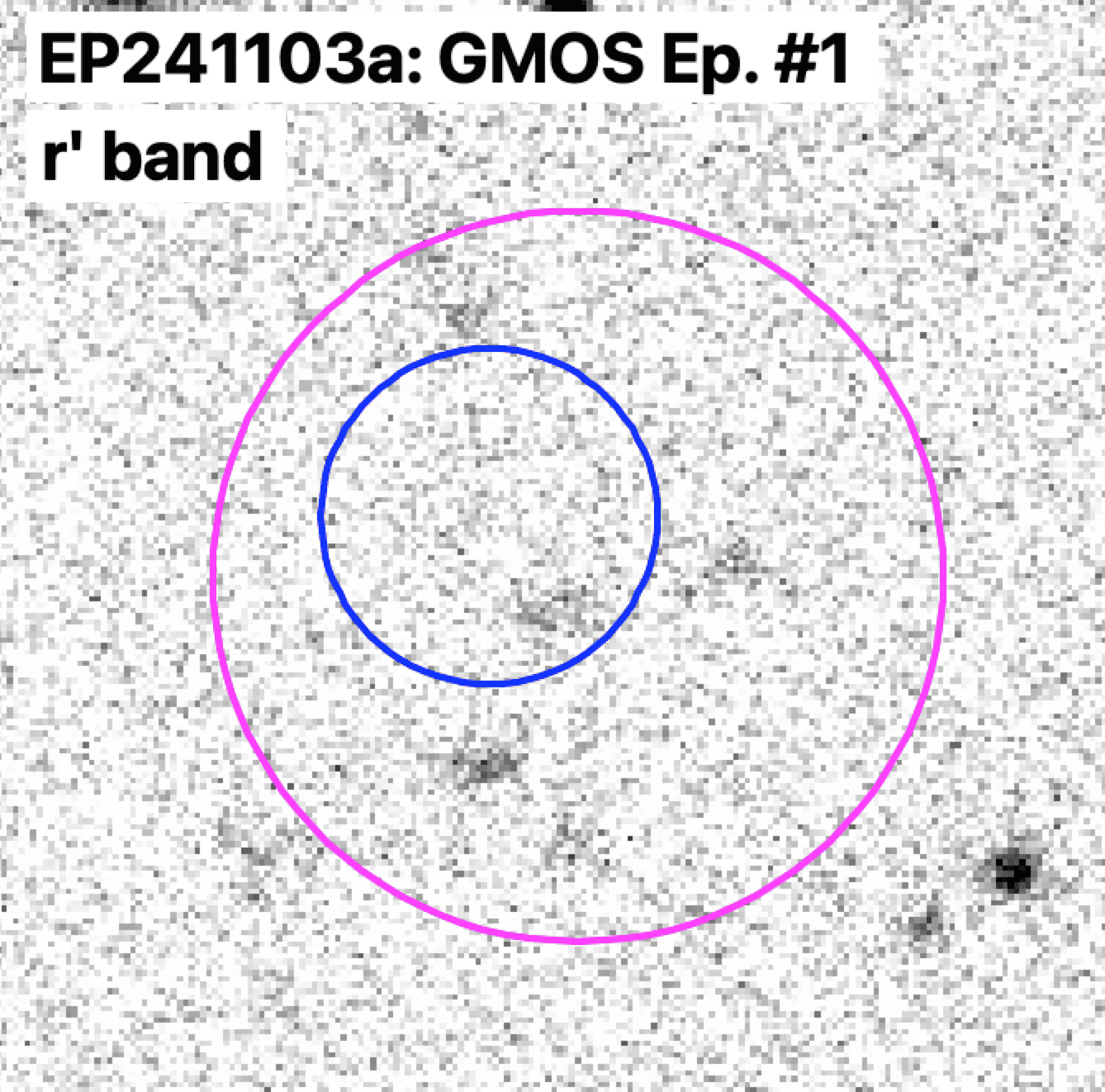}
    \includegraphics[width=5truecm]{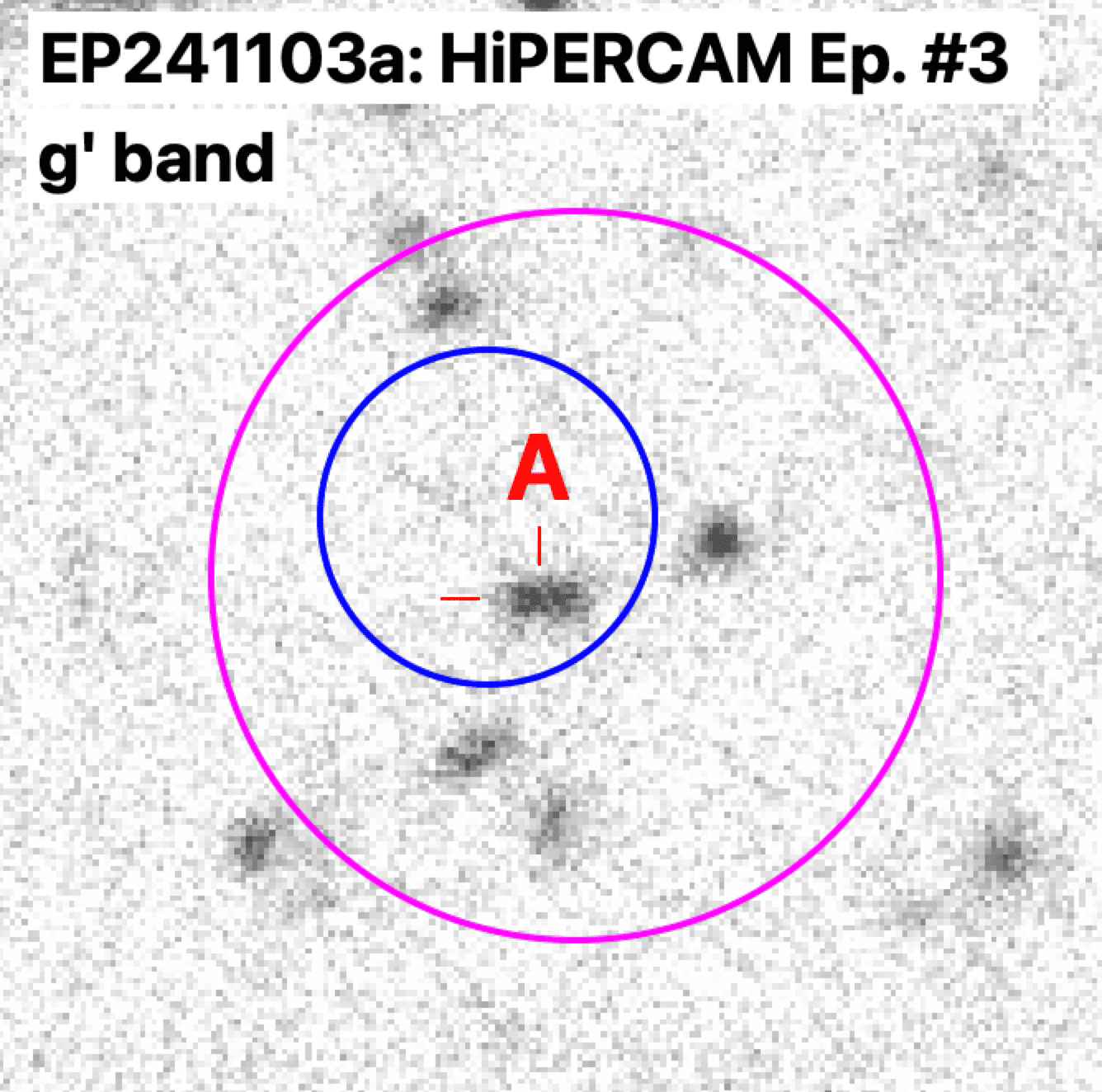}
    \includegraphics[width=5truecm]{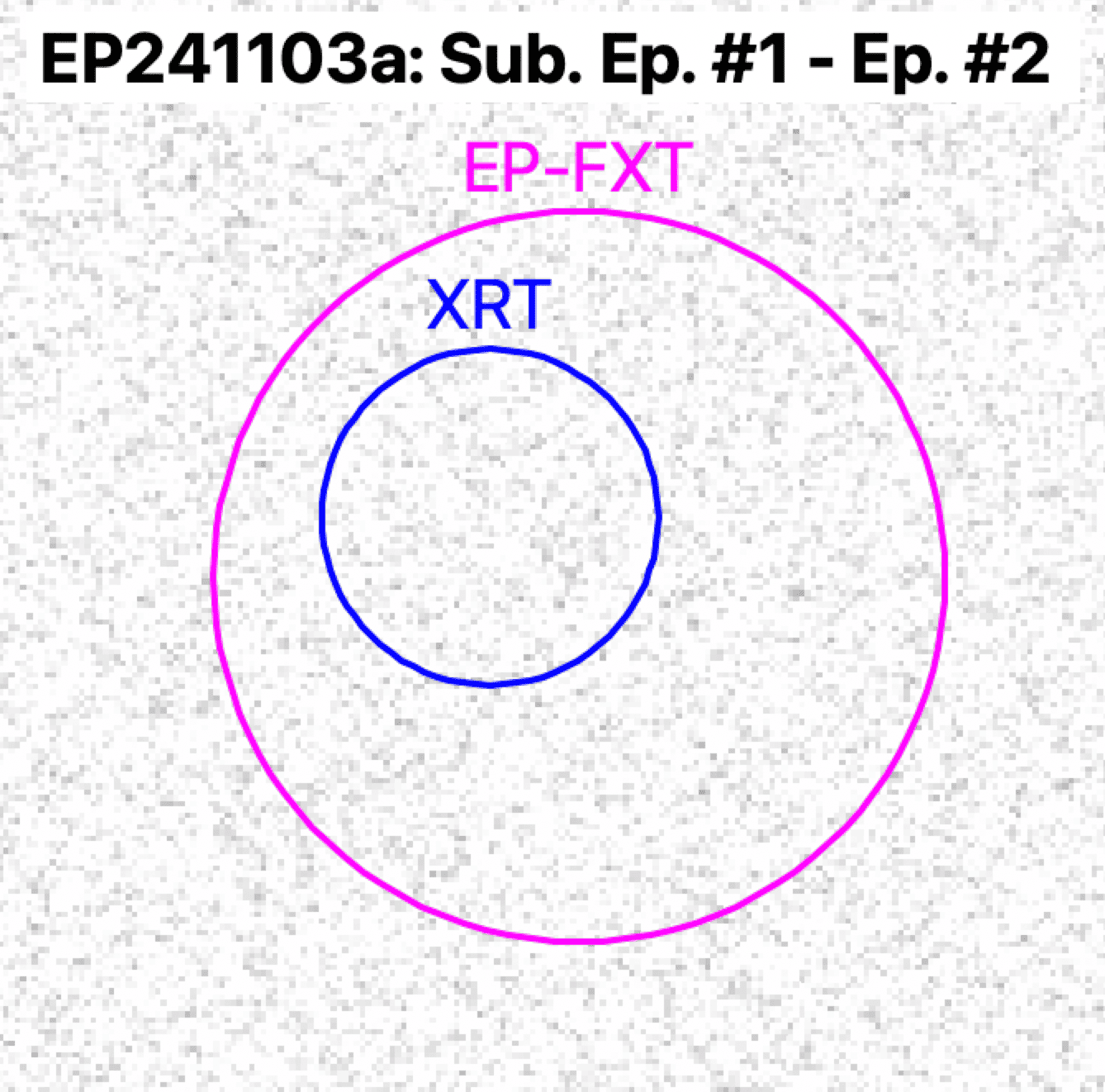}
    \includegraphics[width=5truecm]{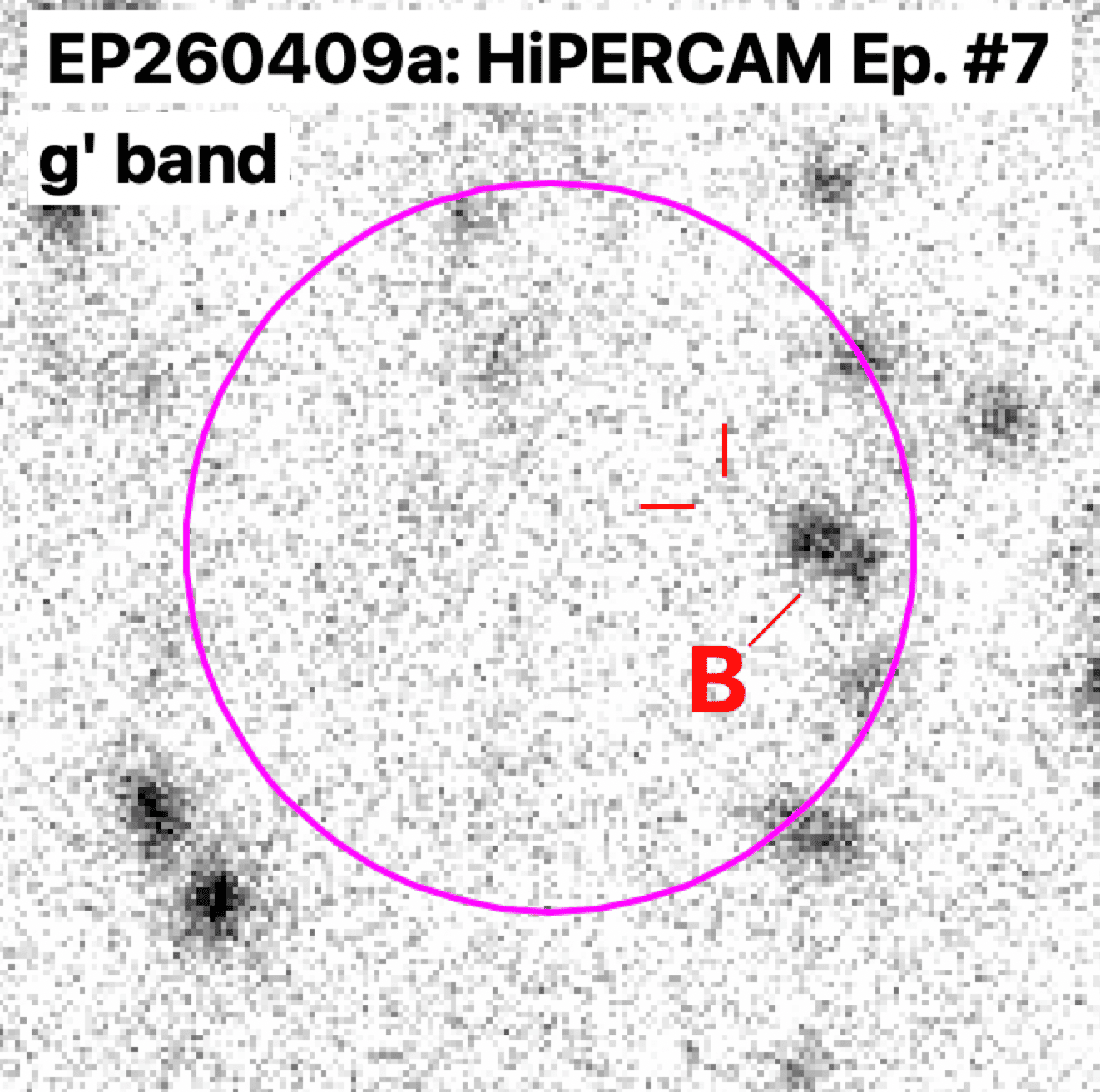}
    \includegraphics[width=5truecm]{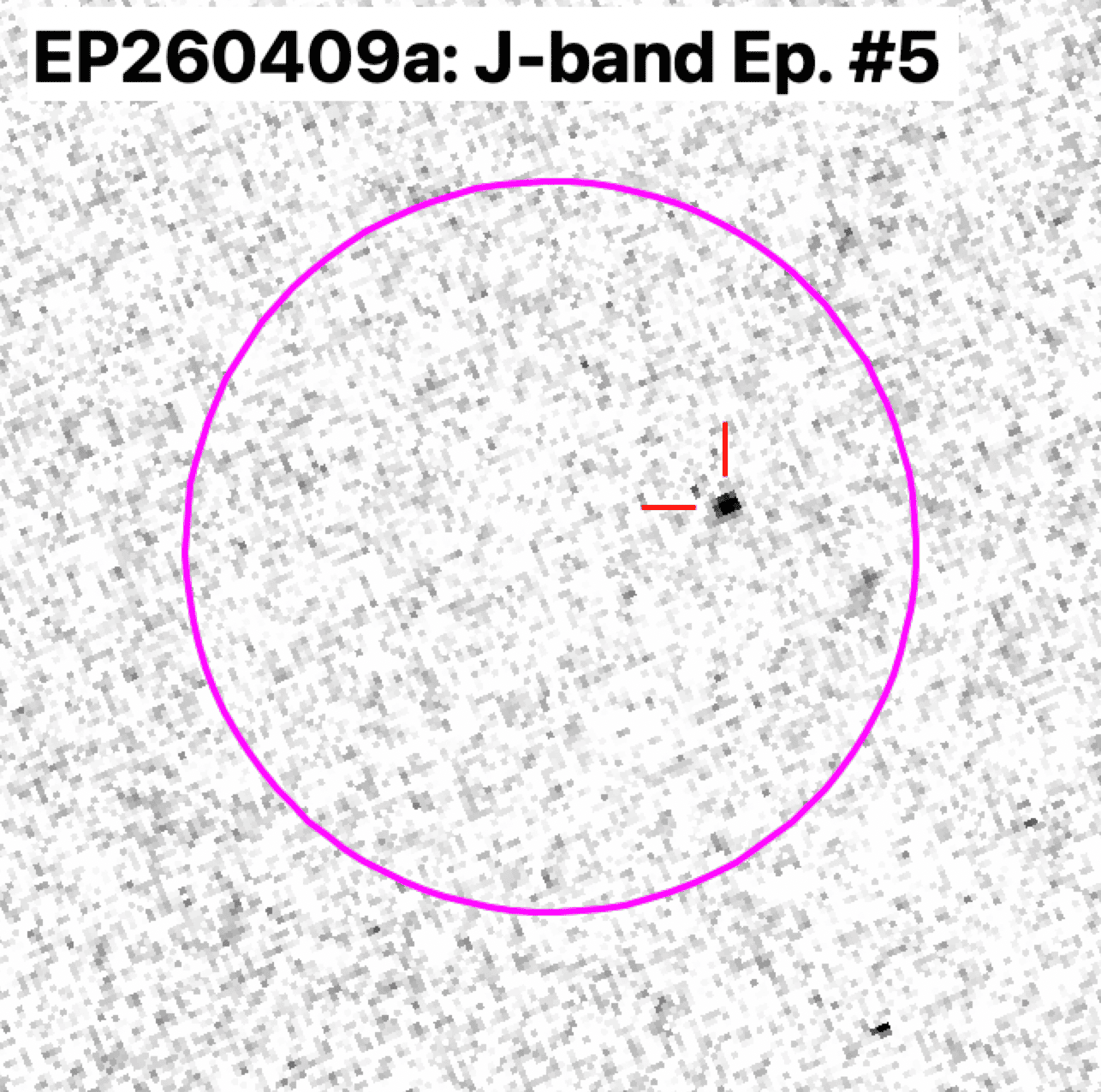}
    \includegraphics[width=5truecm]{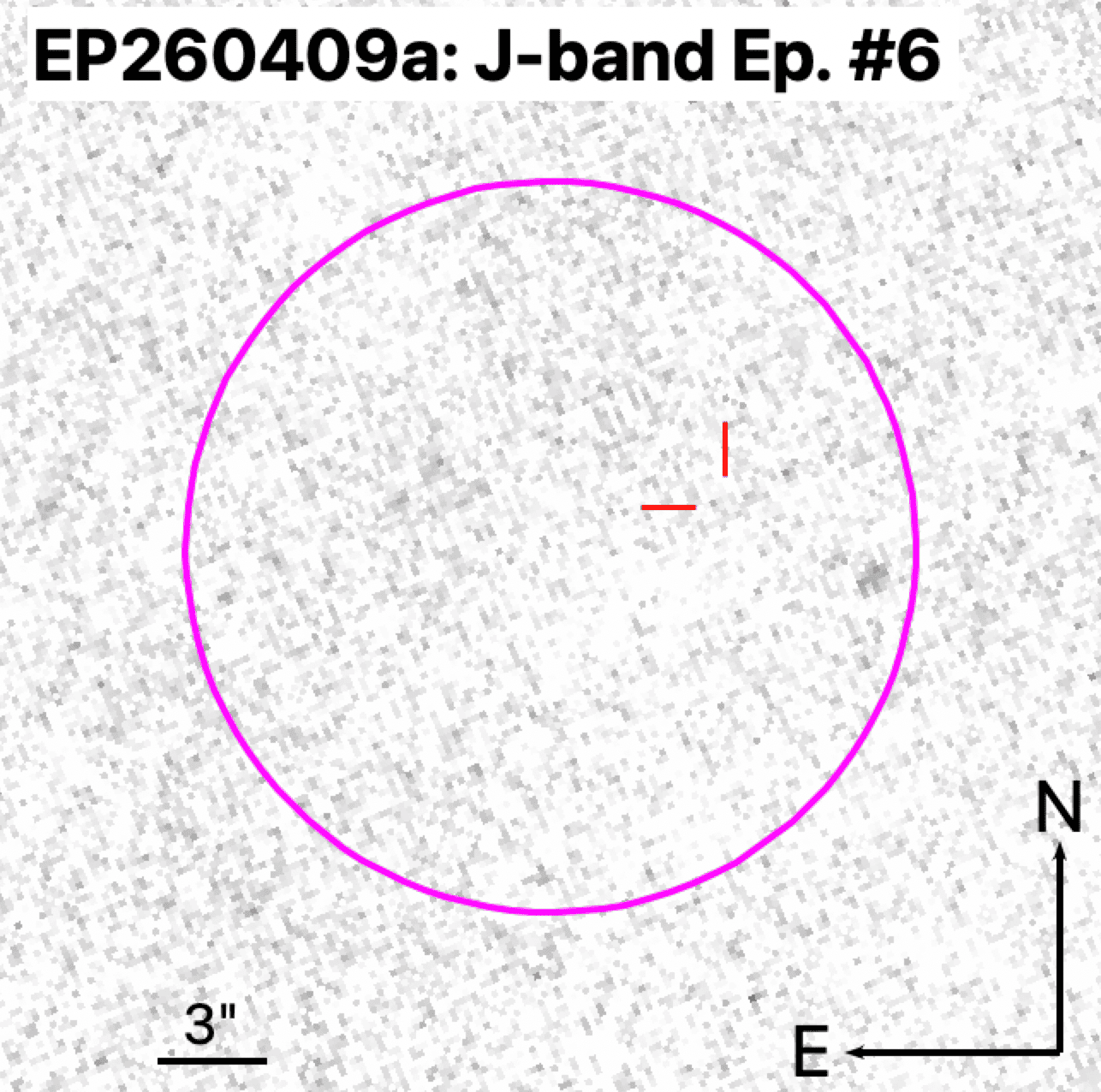}
    \caption{Optical finder charts of \targetone{} (top row) and \targettwo{} (bottom row). Circles denote the uncertainty on the X-ray source position for EP-FXT (magenta) and \textit{Swift}-XRT (blue). The dimensions and orientation are the same in the six panels, as indicated in the bottom-right panel. \textit{\targetone{}}: \textit{Left}: The epoch \#1 GMOS $r^\prime$-band image. \textit{Centre}: The epoch \#3  HiPERCAM $g^\prime$-band image. The host galaxy candidate inside the \textit{Swift}-XRT uncertainty region is labelled as A. \textit{Right}: The image resulting from subtraction of the epochs \#1 and \#2 GMOS $r^\prime$-band images. \textit{\targettwo{}}: The position of the near-infrared counterpart is marked with two red lines in the three panels. \textit{Left}: The epoch \#7 HiPERCAM $g^\prime$-band image. The extended object near the counterpart is labelled as B. \textit{Centre and right}: J-band images from epochs \#5 and \#6. We note that the counterpart is detected in epoch \#5 but not in epoch \#6.}
    \label{figFindingCharts}
\end{figure*}

We obtained a third epoch of \targetone{} using the quintuple-beam camera HiPERCAM \citep{Dhillon2021}, attached to the Folded Cassegrain focus of the 10.4-m Gran Telescopio Canarias (GTC), La Palma, Spain. The instrument acquires images simultaneously in five different filters ($u^\prime$, $g^\prime$, $r^\prime$, $i^\prime$, $z^\prime$). The observation started at \tzerotone{}$ + 47.93$~d and consisted of $80\times30$~s frames. We also secured an observation of \targettwo{} with HiPERCAM, labelled as epoch \#7 in Table \ref{tableObsDetails}, starting at \tzerottwo{}$ + 2.98$~d with a total of $120\times30$~s exposures. Standard data reduction (bias and flatfield correction) was performed using the HiPERCAM pipeline v1.5.6 \citep{Dhillon2018}. We de-fringed the $i^\prime$- and $z^\prime$-band frames by subtracting a scaled fringe map pre-prepared from deep observations. The science frames were stacked for each filter using custom routines executed under \textsc{Python}~3.12.

We observed \targettwo{} using the Sinistro camera on the 1-m node of the Las Cumbres Observatory Global Telescope Network \citep{Brown2013} located at Cerro Tololo, Chile. The observations started at \tzerottwo{}$ + 2.35$~h and consisted of $6 \times 300$~s exposures in each of the $r^\prime$ and $z^\prime$ bands (see Table~\ref{tableObsDetails}). We stacked the individual automatically reduced frames to obtain final stacked images.

We obtained two epochs of \targettwo{} in the near-infrared using the FLAMINGOS2 (F2) instrument \citep{Eikenberry2004} in imaging mode, mounted on GS. The observations were performed in the $J$ band, starting at \tzerottwo{}$ + 6.60$~h and \tzerottwo{}$ + 49.95$~h with exposures of $8 \times 40$ and $16 \times 40$~s, respectively. We reduced the data using the standard imaging steps in the DRAGONS pipeline v4.1.0 \citep{Labrie2023,Simpson2023}.


\section{Analysis and results}
\label{SecResults}

\subsection{X-rays}
\label{secResultsXrays}
For the EP-WXT spectrum (0.5--4~keV) of \targetone{}, we fit an absorbed power-law model in which the line-of-sight Galactic $N_{\mathrm{H}}$ was fixed at $6.3 \times 10^{20}$~cm$^{-2}$. The best-fit power-law index is $\Gamma_{\mathrm{WXT}}^{1103a}=0.9\pm0.3$. For the first two EP-FXT \targetone{} observations we also fit an absorbed power-law model. The photon index does not change significantly between these two epochs ($\Gamma_{\mathrm{FXT,E1}}^{\mathrm{1103a}} = 1.9\pm0.3$ and $\Gamma_{\mathrm{FXT,E2}}^{\mathrm{1103a}}=2.1\pm0.5$, respectively). The source is not detected in the third EP-FXT observation. For the \textit{Swift}-XRT spectrum we obtain a photon index of $\Gamma_{\mathrm{XRT}}^{\mathrm{1103a}} = 1.8^{+3.0}_{-1.3}$. 

\begin{table*}
	\centering
	\caption{Journal of optical and near-infrared observations. We note that the observations in all GTC/HiPERCAM filters are taken simultaneously.}
	\begin{threeparttable}
	\begin{tabular}{c c c c c c c c }
		\hline
		Epoch & Source & Start date (UTC) & Time\tnote{(a)} & Instrument & Band & Exp. time [s] & AB mag
		\\
		\hline
		\hline
		\#1 & \targetone{} & 2024-11-03 06:20:03 & 5.01~h & GMOS & $r^\prime$ & 480 & > 25.0 \\
		\hline
		\multirow{2}{*}{\#2} & \multirow{2}{*}{\targetone{}} & 2024-11-04 03:21:15 & 26.10~h & \multirow{2}{*}{GMOS} & $r^\prime$ & 1200 & > 25.2 \\
		   &  & 2024-11-04 03:47:15 & 26.59~h &  & $z^\prime$ & 1080 & > 24.5 \\
		\hline
		\#3 & \targetone{} & 2024-12-20 21:05:39 & 47.93~d & HiPERCAM & $u^\prime$, $g^\prime$, $r^\prime$, $i^\prime$, $z^\prime$ & 2400 & > 24.7, > 25.3, > 25.3, > 24.7, > 24.2 \\
        \hline
		\multirow{2}{*}{\#4} & \multirow{2}{*}{\targettwo{}} & 2026-04-09 02:37:57 & 2.60~h & \multirow{2}{*}{Sinistro} & $r^\prime$ & 1800 & > 23.9 \\
		   &  & 2026-04-09 03:33:17 & 3.52~h &  & $z^\prime$ & 1800 & > 22.3 \\
		\hline
		\#5 & \targettwo{} & 2026-04-09 06:52:40 & 6.64~h & FLAMINGOS2 & $J$ & 320 & $22.00 \pm 0.10$ \\
		\hline
		\#6 & \targettwo{} & 2026-04-11 02:13:37 & 50.04~h & FLAMINGOS2 & $J$ & 640 & > 23.4 \\
		\hline
		\#7 & \targettwo{} & 2026-04-11 23:23:40 & 2.98~d & HiPERCAM & $u^\prime$, $g^\prime$, $r^\prime$, $i^\prime$, $z^\prime$ & 3600 & --$^{\mathrm{(b)}}$, > 26.0, > 25.2, > 24.2, > 23.2 \\
        \hline
	\end{tabular}
    \begin{tablenotes}
        \item[(a)] Time to mid-observation after \tzerotone{} (epochs \#1--3) and \tzerottwo{} (epochs \#4--7). 
        \item[(b)] The limiting magnitude cannot be determined due to a lack of $u^{\prime}$-band reference stars in the HiPERCAM field of view.
    \end{tablenotes}
	\label{tableObsDetails}
	\end{threeparttable}
\end{table*}

For \targettwo{} the best-fit spectral model yields $\Gamma^{0409a}_{\mathrm{WXT}}=0.4\pm0.7$. Both EP-FXT observations detect \targettwo{} and we obtain photon index values of $\Gamma_{\mathrm{FXT,E1}}^{\mathrm{0409a}} = 1.6 \pm 0.2$ and $\Gamma_{\mathrm{FXT,E2}}^{\mathrm{0409a}} = 1.5 \pm 0.8$ in the 0.5--10~keV band. Additional information is provided in the online supplementary material.

The observations of both targets are consistent with a constant X-ray spectral power law with $\Gamma^{\mathrm{1103a}} = 2.0\pm0.4$ and $\Gamma^{\mathrm{0409a}} = 1.6\pm0.2$, respectively. The flux also decays as a power-law ($F_{\nu} \propto t^{-\alpha}$), as typically also seen in GRB afterglows \citep{Sari1998}. We obtain $\alpha^{\mathrm{1103a}} = 0.34\pm0.09$ and $\alpha^{\mathrm{0409a}} = 0.30\pm0.07$. We use these EP-FXT fluxes to interpolate the X-ray flux to the times of our optical observations. We note that the derived $\alpha^{\mathrm{1103a}}$ value is consistent with both the X-ray flux measured at the Swift-XRT observation, as well as the flux upper limit of the non-detection during the third EP-FXT epoch.

\subsection{Optical and near-infrared}
No new source in the optical images of \targetone{} and \targettwo{} was found within the EP-FXT or \textit{Swift}-XRT error circle compared to archival data from the deep optical survey DESI Legacy Imaging Survey (\citealt{Dey2019}; Legacy hereafter). We obtain limiting magnitudes for each image using the Pan-STARRS survey ($g^\prime$, $r^\prime$, $i^\prime$ and $z^\prime$ bands; \citealt{Magnier2020}) and The Sloan Digital Sky Survey DR8 ($u^\prime$ band; \citealt{Aihara2011}) for photometric calibration. We were unable to calibrate the $u^\prime$-band image of \targettwo{} (epoch \#7) due to the lack of reference stars in the HiPERCAM field of view. To obtain 3$\sigma$ limiting magnitudes, we mask the objects present in the image, fit the Point Spread Function (PSF) using EPSFBuilder from the photutils package \citep{Price-Whelan2022}, and calculate the median absolute deviation (MAD). We converted the MAD to a Gaussian-equivalent standard deviation ($\sigma = 1.4826~\mathrm{MAD}$) and corrected the 90\% encircled-energy flux to total flux before deriving the $3\sigma$ magnitude limits, which are provided in Table~\ref{tableObsDetails}.

\label{secObservationsSubtraction}
For \targetone{}, we performed image subtraction between the different epochs and against images from Legacy using the ZOGY algorithm \citep{Zackay2016}, as implemented in PyZOGY \citep{Guevel2021} for each filter of each epoch. Background sky subtraction and other image operations were performed using \textsc{Astropy} tools \citep{Bradley2025}. For the GMOS data, we also used the epoch \#2 $r^\prime$-band image as a template for subtraction from epoch \#1. Fig.~\ref{figFindingCharts} (top-right panel) shows the subtraction of these two images, in which no new source is detected inside the EP-FXT or \textit{Swift}-XRT error circle. We also used the epoch \#3 HiPERCAM $r^\prime$- and $z^\prime$-band images as templates to subtract from the matching-filter epoch \#1 and \#2 images. None of these subtractions revealed a new source in the EP-FXT or \textit{Swift}-XRT uncertainty regions.

An uncatalogued source is detected in the first $J$-band F2 observation of \targettwo{} \citep{gcnQuirolaVasquez2026} at the coordinates $\mathrm{RA_{J2000}} = 11^{\mathrm{h}}31^{\mathrm{m}}37.98^{\mathrm{s}}$, $\mathrm{Dec_{J2000}} = -9^{\circ}20^{\prime}43.3^{\prime\prime}$, with an uncertainty radius of $0.3^{\prime\prime}$. We measure $J = 22.00 \pm 0.10$ (epoch \#5, see Table~\ref{tableObsDetails}) using reference stars from The Two Micron All Sky Survey (2MASS; \citealt{Skrutskie2006}). This source was not detected in the second F2 observation, with a limiting magnitude of $J > 23.4$ (see Fig.~\ref{figFindingCharts}).

\subsection{Spectral Energy Distribution}
\label{secResultsSED}

\begin{figure}
    \centering
    \includegraphics[width=8truecm]{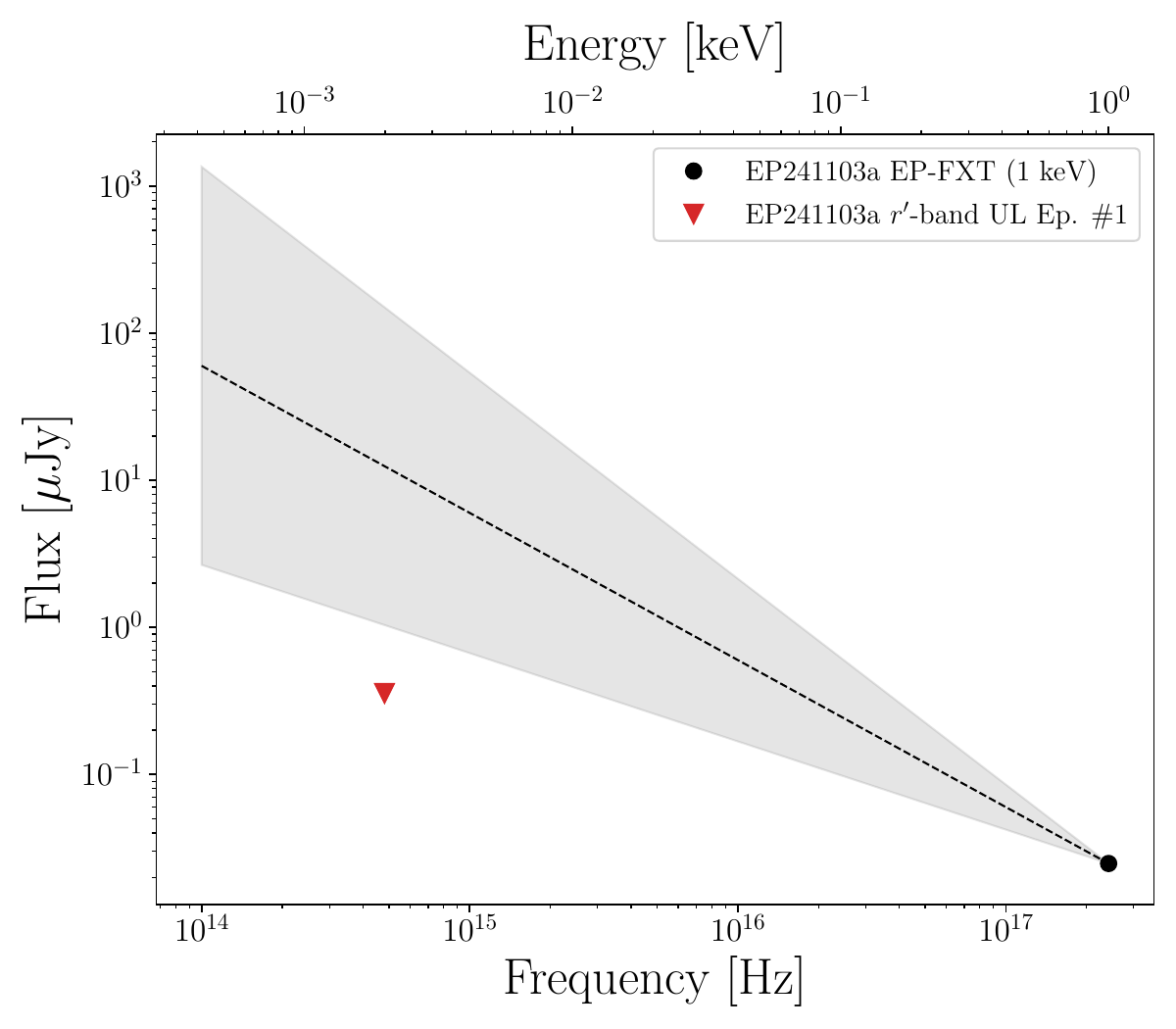}
    \caption{Spectral Energy Distribution (SED) for \targetone{} at the time of our first optical observation ($\sim 5$~h after the EP-WXT trigger). We show the $r^\prime$-band 3$\sigma$ upper limit (red triangle) and the monochromatic (1 keV) X-ray flux derived from the interpolation of the X-ray light curve to the time of the optical observation (black circle). The dashed-black line indicates the extrapolated X-ray flux to optical wavelengths (see Sec.~\ref{secResultsSED}) and the cone-shaped region (grey) denotes the uncertainty range obtained by extrapolating the flux using the uncertainty of the best-fit power-law index. The $r^\prime$-band upper limit lies well below the flux predicted by the extrapolation.}
    \label{figSED}
\end{figure}

Figure~\ref{figSED} shows the broad-band spectral energy distribution (SED) of \targetone{} at the observation time of GMOS epoch \#1 ($\sim 5$~h after \tzerotone{}). The X-ray flux at 1~keV and the extrapolation to the optical wavelength are shown, including the uncertainty from the best-fit power law to the X-ray spectrum as a cone-shaped region. The epoch \#1 $r^{\prime}$-band upper limit is indicated by the red triangle; it lies at approximately two orders of magnitude lower flux than predicted.

\begin{figure}
    \centering
    \includegraphics[width=8truecm]{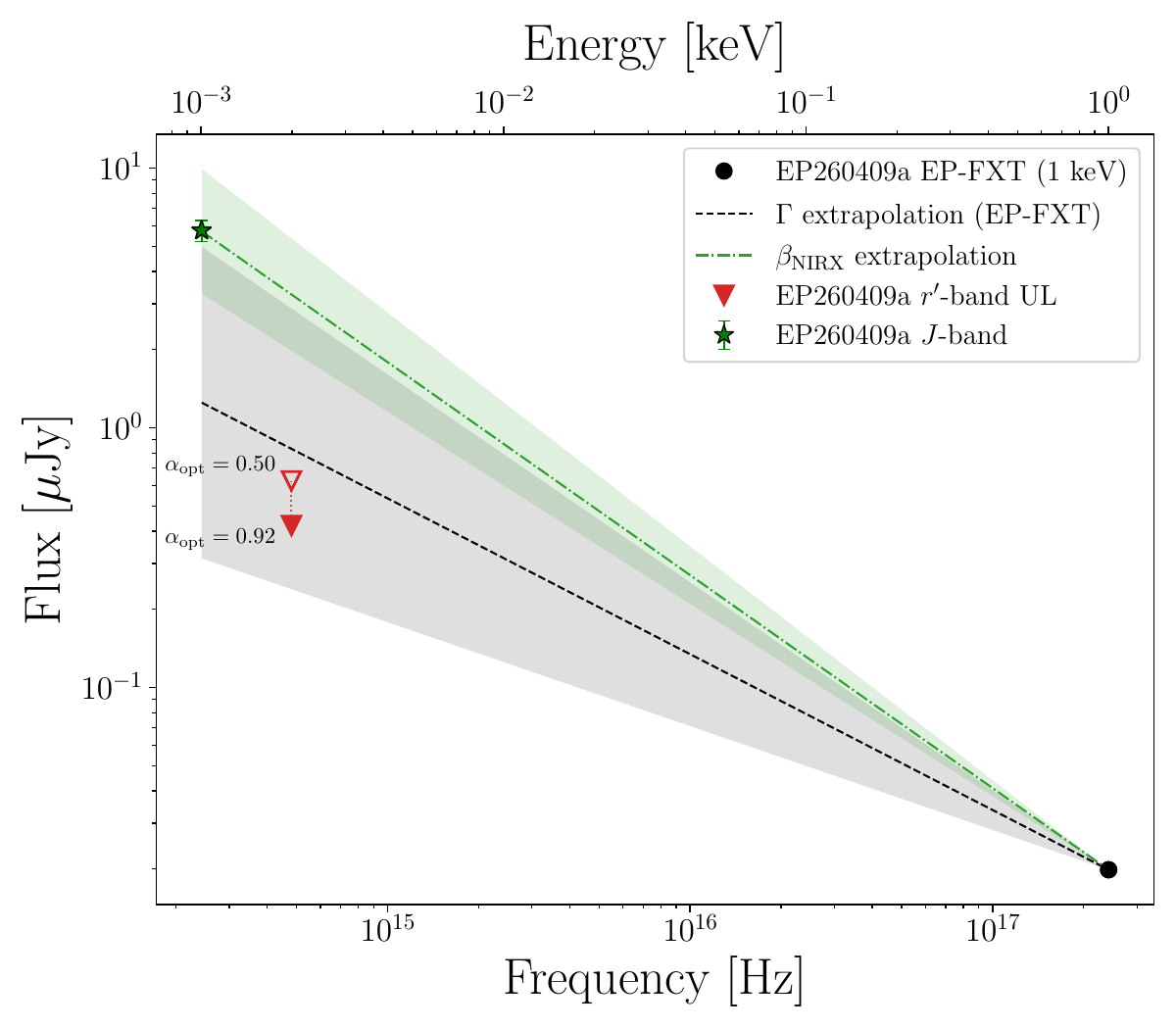}
    \caption{SED for \targettwo{} at the time of the $J$-band epoch \#5 (\tzerottwo{}$+6.64$~h). The $r^\prime$-band extrapolations using $\alpha=0.50$ and $\alpha=0.92$ ($F_{\nu} \propto t^{-\alpha}$) are shown as open and filled red triangles, respectively. The $J$-band photometry is shown as a green star and the interpolated monochromatic flux (1~keV) is shown as a black dot. The dashed-black and dot-dashed-green lines indicate the extrapolated X-ray power law to the optical wavelengths with indices $\Gamma$ and ($1+\beta_{\mathrm{NIRX}}$), respectively. The extrapolated flux using the photon index uncertainties are marked as grey and green cones.}
    \label{figSED0409a}
\end{figure}

We calculate the upper limit on the optical-to-X-ray spectral index ($\beta_{\mathrm{OX}}$) and compare it with the X-ray spectral index ($\beta_{\mathrm{X}}$). We follow the criteria typically used to define a GRB as optically dark: $\beta_{\mathrm{OX}} < 0.5$ \citep{Jakobsson2004} and $\beta_{\mathrm{OX}} < \beta_{\mathrm{X}} - 0.5$ \citep{vanderHorst2009}. The position of \targetone{} in the $\beta_{\mathrm{OX}} - \beta_{\mathrm{X}}$ plane falls within the region of dark GRBs, with the upper limit $\beta_{\mathrm{OX}} < 0.43$ and $\beta_{\mathrm{X}} = 1.0 \pm 0.4$, fulfilling both of the aforementioned criteria and supporting the dark nature of \targetone{}. For \targettwo{}, we obtain an upper limit of $\beta_{\mathrm{OX}} < 0.58$ and $\beta_{\mathrm{X}} = 0.6 \pm 0.2$. We note that the $\beta_{\mathrm{OX}}$ and $\beta_{\mathrm{X}}$ constraints do not formally satisfy these dark criteria. Assuming an achromatic power-law temporal decay in the near-infrared and in the optical, we calculate the near-infrared-to-X-ray spectral index ($\beta_{\mathrm{NIRX}}$) using the $J$-band counterpart photometry from epoch \#5, obtaining $\beta_{\mathrm{NIRX}}=0.82\pm0.12$. We extrapolated the $r^{\prime}$-band limiting magnitude to the time of $J$-band epoch \#5 using temporal decay indices ($F_{\nu} \propto t^{-\alpha}$) of $\alpha = 0.92$ \citep{Ror2025} and a conservative shallower value of $\alpha = 0.50$. Fig. \ref{figSED0409a} shows the SED of \targettwo{}, together with the $J$-band counterpart and the $r^{\prime}$-band extrapolations at the common time of the $J$-band epoch \#5. The X-ray (1~keV) flux and the extrapolations implied by the EP-FXT $\Gamma$ and by $\beta_{\mathrm{NIRX}}$ to near-infrared wavelengths are also shown. The contrast between the upper limit on $\beta_{\mathrm{OX}}$ and $\beta_{\mathrm{NIRX}}$, together with the common-epoch (conservatively assuming $\alpha_{\mathrm{opt}}=0.5$) colour $r^{\prime}-J>2.41$, suggests a very red counterpart. Such a red counterpart can either be explained by extinction or a high-redshift event.

\subsection{Host galaxy candidates}

We calculate the chance alignment probability ($P_{\mathrm{ch}}$; \citealt{Bloom2002}) with the coordinates obtained from \textit{Swift}-XRT for the extended source inside the \targetone{} \textit{Swift}-XRT error circle (labelled as A in Fig. \ref{figFindingCharts}), and for other sources present in the \targetone{} EP-FXT error circle. For object A, we obtained $P_{\mathrm{ch}}(\mathrm{A})=0.032$. In contrast, the $P_{\mathrm{ch}}$ we calculated for the other objects present inside the EP-FXT error circle have values ranging between 40\%--70\%. We perform photometry on object A using the epoch \#3 images. The multi-filter observation from HiPERCAM allowed us to fit the photometric measurements with Prospector \citep{Johnson2021}. We obtain a photometric redshift $z^{A}_{\mathrm{ph}} = 2.50^{+0.17}_{-0.71}$. 

Following the same procedure, we calculate $P_{\mathrm{ch}}$ using the angular separation between the \targettwo{} J-band counterpart and the centre of object B (Fig.~\ref{figFindingCharts}), obtaining $P_{\mathrm{ch}}(\mathrm{B})=0.18$. We also used the photometry obtained from the \targettwo{} HiPERCAM observation (epoch \#7) to estimate $z^{B}_{\mathrm{ph}} = 0.61^{+0.06}_{-0.10}$ with Prospector.

\section{Discussion}
\label{secDiscussion}

Comparing dark GRBs with dark FXTs, the discovery of dark GRBs is by definition made using $\gamma$- or hard X-rays, where photoelectric absorption is negligible. Therefore, setting aside potential differences in the absolute magnitudes of the various bursts, optical darkness typically implies either high dust extinction or a redshift high enough for the Ly-break to be redshifted beyond the optical band in the observer frame \citep{Perley2009,Melandri2012}, with the latter being a minority ($\sim 14$\%; \citealt{Chrimes2019}). The darkness of dark GRBs has been found to be tightly correlated with the value of $N_{\mathrm{H}}$, with most of them lying in the range $21.5 \lesssim \mathrm{log}[N_{\mathrm{H}}/\mathrm{cm}^{-2}] \lesssim 22.7$ \citep{Zheng2009,Campana2012}. The absence or faintness of the optical counterpart means that for these events the redshift is more difficult to obtain from the afterglow. Nevertheless, for some, a likely redshift can be obtained if it is possible to measure the redshift of an associated host galaxy. Since these hosts typically have low masses and low metallicities \citep{Perley2016a}, they are faint and, therefore, dark GRBs with an inferred redshift are usually found at low redshift (e.g.~\citealt{Chrimes2019}). In contrast, low-redshift dark FXTs are probably rare. Currently, FXTs are mainly discovered in the 0.5--4~keV X-ray band of EP-WXT. Although the effective area drops quickly above $\sim1$~keV, making it especially sensitive to photons in the energy range 0.5--1~keV \citep{Yuan2022}. Such photons are strongly affected by photoelectric absorption in atoms in neutral gas ($N_{\mathrm{H}}$). Any intrinsic absorption will therefore affect the observed X-ray flux for low redshift events. The cross-section for photoelectric absorption for most elements drops rapidly with X-ray photon energy. Therefore, for FXTs at redshifts $z$~$\approxgt 2-2.5$ the column density of local absorbing material needs to be extremely high in order to still affect the observed X-ray emission.

\begin{figure}
    \centering
    \includegraphics[width=8.5truecm]{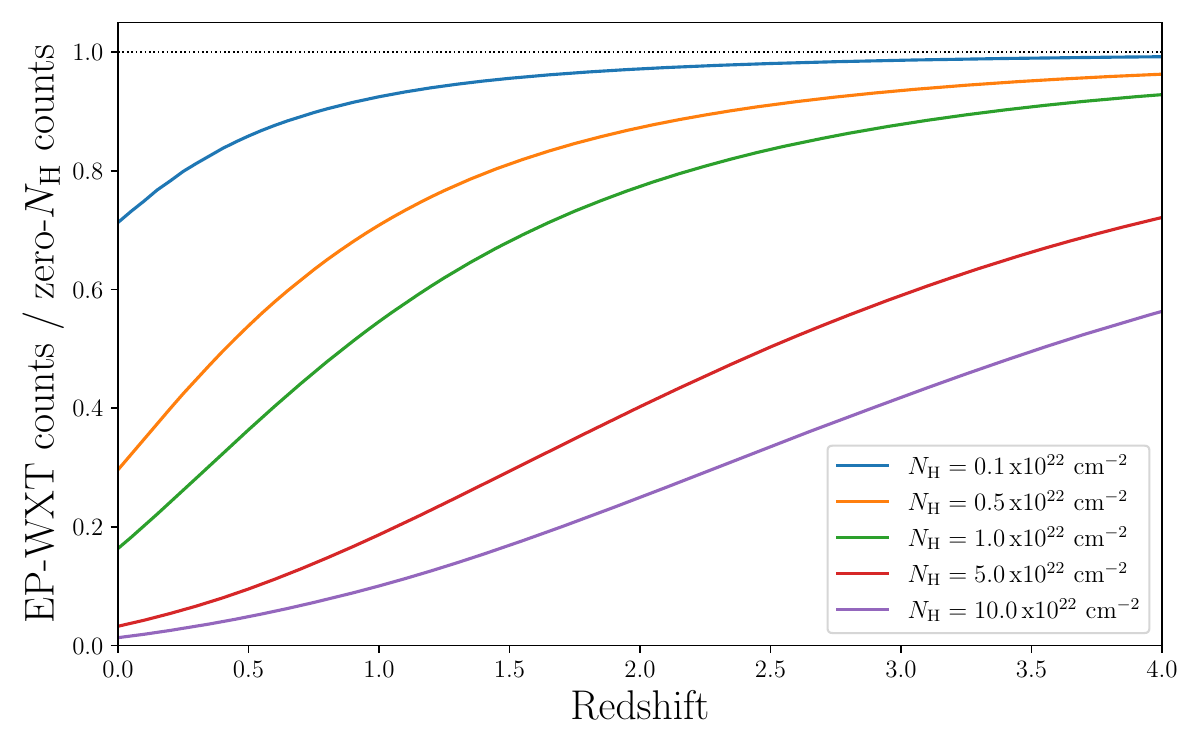}
    \caption{Simulated ratio of counts received by the EP-WXT (0.5--4~keV) from a source with different $N_\mathrm{H}$ values in the range $10^{21}$~cm$^{-2} \leq N_{\mathrm{H}} \leq 10^{23}$~cm$^{-2}$ to counts from the same source with $N_\mathrm{H} = 0$ as a function of redshift.}
    \label{figEPWXT_ZeroNH}
\end{figure}

To study the detectability by EP of highly absorbed FXTs at different redshifts, we simulate FXT observations with EP-WXT\footnote{The EP-WXT response and ancillary files, available in the Data Availability section, were obtained using the EP-WXT Data Simulator: https://ep.bao.ac.cn/ep/simulator/}. We use the \texttt{fakeit} function from the \texttt{XSPEC} tool implemented in PyXspec \citep{Gordon2021} to simulate an FXT spectrum with the EP-WXT response. We then calculate the ratio between the model-predicted count rate for an absorbed source with different intrinsic $N_{\mathrm{H}}$ in the range $10^{21}$~cm$^{-2} \leq N_{\mathrm{H}} \leq 10^{23}$~cm$^{-2}$ and that obtained with zero intrinsic absorption, for redshifts between $0 \leq z \leq 10$. Fig. \ref{figEPWXT_ZeroNH} shows this ratio as a function of redshift in the range $0 \leq z \leq 4$. The results of these simulations show that, while the ratio is close to one (i.e.~detectability is weakly affected) for low intrinsic $N_{\mathrm{H}}$ values (i.e.~$\sim 10^{21}$~cm$^{-2}$), in moderately and strongly absorbed events (i.e.~$N_{\mathrm{H}} > 10^{22}$~cm$^{-2}$) the predicted EP-WXT count rate is strongly suppressed at low redshift ($z < 1$). Instead, as an example, if the FXT is located at $z \sim 2.5$, the photons received by EP at $\sim 1$~keV were emitted at $\sim 3.5$~keV in the rest frame, requiring $N_\mathrm{H} > 10^{23}$~cm$^{-2}$ to reduce the flux by $\sim 50$\% (\citealt{Morrison1983}; see also Fig. \ref{figEPWXT_ZeroNH}). We note the dependence of $N_{\mathrm{H}}$ on redshift as $N_{\mathrm{H}}(z=0) = N_{\mathrm{H}}(z) / (1 + z)^{a}$, where the scaling index $a$ can vary depending on the cut-off energy and the instrument ($2.34 \lesssim a \lesssim 2.42$; \citealt{Campana2014}); however, this should not strongly affect these broad trends if there is no intervening absorber along the line-of-sight (see also \citealt{Campana2012}). An additional caveat is that the existing correlations between $N_{\mathrm{H}}$ and $A_{V}$ yield different values in the range $2.1 \lesssim N_{\mathrm{H}} / A_{V} [\times 10^{21}~\mathrm{cm}^{-2}~\mathrm{mag}^{-1}] \lesssim 3.4$ \citep{Guver2009,Watson2011,Foight2016}, and can be broken for certain GRBs (e.g. \citealt{Kruhler2011,Covino2013}). It is also worth noting that the prediction of a different redshift distribution function for dark GRBs and dark FXTs contrasts with the similarity between the redshift distributions of the broader GRB and FXT populations, as shown in other works (e.g. \citealt{OConnor2025,Guo2025,Li2026b}). However, this is perhaps not surprising, as that prediction is for the detectability of the dark FXTs and dark GRBs. It is not implying an intrinsically different redshift distribution. Alternatively, the optical darkness of dark FXTs could also be caused by an intrinsically faint optical afterglow \citep{Berger2002}, for instance due to less energy input from the central engine in the forward shock model (e.g. \citealt{Urata2007}) or to a low-density interstellar medium through which the outburst propagates \citep{Melandri2012,Volnova2014}.

Based on the early optical upper limiting magnitude, as well as its position in the $\beta_{\mathrm{OX}} - \beta_{\mathrm{X}}$ plane, we classify \targetone{} as a dark FXT. The chance alignment probability of the only galaxy we detect inside the \textit{Swift}-XRT error circle of \targetone{} ($P_{\mathrm{ch}} \sim 0.032$) is such that we consider the galaxy as a candidate host galaxy (e.g. $P_{\mathrm{ch}} \lesssim 0.1$; \citealt{OConnor2022}). If \targetone{} is indeed at $z = 2.5$, the EP-WXT spectrum modelled with extra intrinsic absorption ($N_{\mathrm{H,int}}$) in addition to the Galactic absorption yields a best-fit photon index $\Gamma = 2.2 \pm 0.8$ and an intrinsic hydrogen column density of $N_\mathrm{H,int} = \left(8 \pm 4\right) \times 10^{22}$~cm$^{-2}$, which is significantly higher than the Galactic $N_{\mathrm{H}}$. In order to statistically compare this model with the one without the intrinsic absorption component, we calculated the Bayesian and Akaike information criteria ($BIC$ and $AIC$) for both models. For the first model (\texttt{phabs*pow}), we obtain $BIC = 93.5$ and $AIC = 88.6$, while for the second (\texttt{phabs*zphabs*pow}, at $z = 2.5$) we get $BIC = 85.7$ and $AIC = 78.3$. Therefore, we conclude that the fit function including the intrinsic host galaxy absorption at $z = 2.5$ is statistically favoured. In the optical, epoch \#1 was obtained at $\sim$\tzerotone{}$+5$~h, showing no new sources inside the EP-FXT and \textit{Swift}-XRT uncertainty regions. At a redshift of $z = 2.5$, the  optical flux received in the $r^{\prime}$ band was emitted at ultraviolet wavelengths in the rest frame ($\sim 1784$~\AA). We transform the $N_\mathrm{H,int}$ into $A_{V}$ using the relation of \cite{Guver2009}, and convert $A_{V}$ into extinction at $\sim 1784$~\AA~assuming the ultraviolet (UV) and far-UV extinction curve from \cite{Cardelli1989} and $R_{V} = 3.1$. We obtain $A_{V} = 36^{+23}_{-19}$ and $A_{\lambda1784\text{\AA}} = 89^{+58}_{-47}$ magnitudes of extinction, respectively. This would suppress the $r^{\prime}$-band flux inferred from the X-ray extrapolation by a factor of $\sim \times 10^{35}$, easily placing it below the observed upper limit. We note that there are additional systematic uncertainties, such as the error in $z_{\mathrm{ph}}$ and the underlying uncertainty on whether the host association based on the low $P_{\mathrm{ch}}$ is correct. As a matter of fact, \citealt{Jonker2026} reported a $z_{\mathrm{ph}} = 0.082$ based on a $P_{\mathrm{ch}} < 0.005$~ from a host candidate for EP250207b (see also \citealt{Becerra2026}); however, deep late time space-based observations revealed a galaxy at higher redshift underneath \citep{vanHoof2026c}. These systematic and large statistical uncertainties notwithstanding, the derived extinction would readily explain the non-detections in the optical during epochs \#1 and \#2, while the X-rays would still be detected.

For \targettwo{}, the uncatalogued object detected in the first $J$-band observation (epoch \#5), which is not detected in the $J$-band epoch \#6, allows us to conclude that this is the near-infrared counterpart of this FXT. The relatively high $P_{\mathrm{ch}}$ for B (see Fig.~\ref{figFindingCharts}) implies that the probability that this galaxy is the host is low. At $z_{\mathrm{ph}}^{\mathrm{B}} = 0.6$, the projected distance to the centre of the galaxy would be 22~kpc. Although a compact-object merger origin could be geometrically plausible for this separation (e.g.~\citealt{Fong2022}), it is large. \citet{OConnor2022} found a median projected offset for short GRBs of 5.6~kpc and, together with the absolute magnitude of $M_{J}=-20.7$ derived from the photometry at $\sim 7$~h after \tzerottwo{}, it would disfavour an AT2017gfo-like kilonova (KN) origin \citep{Valenti2017}. We note that $z_{\mathrm{ph}}^{\mathrm{B}}$ could be incorrect; however, the redshift would have to be much lower for the $J$-band photometry to be explained as KN emission ($z\lesssim0.1$ for $M_{J} \gtrsim -16.4$; \citealt{Ascenzi2019}), and, in that case, the host would have to be an intrinsically very faint dwarf galaxy. Furthermore, the $J$-band emission should have increased at epoch \#6 instead of disappearing \citep{Metzger2020}, making the KN origin unlikely. A source with high intrinsic absorption remains a possible interpretation, but the large projected offset from the host argues against this scenario \citep{Lyman2017}. If \targettwo{} is not associated with B, it could be a high-redshift FXT. This would naturally explain the early optical non-detections, which would be due to suppression by the Ly-$\alpha$ forest or the Lyman break ($z \gtrsim 4.1$ for $r^{\prime}$ band), while the $J$-band would remain unaffected. The lack of a host in the deep epoch \#7 observations is also consistent with a faint or distant galaxy.

\section{Conclusions}
\label{secConclusions}

The discovery energy bands of dark FXTs and dark GRBs can introduce a different selection effect. While dark GRBs with known redshift are biased towards low redshift values, dark FXTs with high intrinsic absorption can hardly be detected at low redshift because the X-ray absorption affects the band in which they are discovered, hindering their discovery in the first place. This effect does not occur at intermediate (and higher) redshift, producing a selection effect in dark FXTs towards these redshifts and this differs from that affecting dark GRBs. We presented observations of the optically dark FXTs \targetone{} and \targettwo{}, which led us to conclude that the most plausible scenario for \targetone{} is that the source is located at a redshift of $z \sim 2.5$, while \targettwo{} is located at a high redshift ($z \gtrsim$ 4). Under these scenarios, the optical darkness of \targetone{} is due to high intrinsic absorption, while that of \targettwo{} is caused by the optical suppression due to the Ly-$\alpha$ forest or the Lyman break. The detection of these dark FXTs supports and readily fits into our general prediction that dark FXTs and dark GRBs have different selection functions, and highlights the importance of the dark FXTs for studying and discovering populations at moderate and high redshifts.

\section*{Acknowledgements}

JSS, PGJ, JQV, APCvH, MER, and JNDvD are supported by the European Union (ERC, StarStruck, 101095973; PI Jonker). Views and opinions expressed are however those of the author(s) only and do not necessarily reflect those of the European Union or the European Research Council. Neither the European Union nor the granting authority can be held responsible for them. This work is based on data obtained with Einstein Probe, a space mission supported by the Strategic Priority Program on Space Science of the Chinese Academy of Sciences, in collaboration with ESA, MPE, and CNES.  MER is supported by the ”la Caixa” Foundation Junior Leader Fellowship (ID 100010434). The fellowship code is LCF/BQ/PI25/12100030. This work was also partly supported by the Spanish program Unidad de Excelencia Maria de Maeztu CEX2020-001058-M, financed by MCIN/AEI/10.13039/501100011033, and by the MaX-CSIC Excellence Award MaX4-SOMMA-ICE. DMS acknowledges support through the Ramon y Cajal grant RYC2023-044941, funded by MCIU/AEI/10.13039/501100011033 and FSE+. FEB acknowledges support from ANID-Chile BASAL CATA FB210003 and FONDECYT Regular 1241005. This work made use of data supplied by the UK Swift Science Data Centre at the University of Leicester. This work is partly based on observations obtained at the international Gemini Observatory, a program of NSF NOIRLab, which is managed by the Association of Universities for Research in Astronomy (AURA) under a cooperative agreement with the U.S. National Science Foundation on behalf of the Gemini Observatory partnership: the U.S. National Science Foundation (United States), National Research Council (Canada), Agencia Nacional de Investigaci\'{o}n y Desarrollo (Chile), Ministerio de Ciencia, Tecnolog\'{i}a e Innovaci\'{o}n (Argentina), Minist\'{e}rio da Ci\^{e}ncia, Tecnologia, Inova\c{c}\~{o}es e Comunica\c{c}\~{o}es (Brazil), and Korea Astronomy and Space Science Institute (Republic of Korea), under program GS-2024B-Q-131. This work is partly based on observations made with the Gran Telescopio Canarias (GTC), installed at the Spanish Observatorio del Roque de los Muchachos of the Instituto de Astrofísica de Canarias, on the island of La Palma, under program GTC1-24ITP. These data were obtained with the instrument HiPERCAM, built by the Universities of Sheffield, Warwick and Durham, the UK Astronomy Technology Centre, and the Instituto de Astrofísica de Canarias. Development of HiPERCAM was funded by the European Research Council, and its operations and enhancements by the Science and Technology Facilities Council. This work makes use of observations from the Las Cumbres Observatory global telescope network. This publication makes use of data products from the Two Micron All Sky Survey, which is a joint project of the University of Massachusetts and the Infrared Processing and Analysis Center/California Institute of Technology, funded by the National Aeronautics and Space Administration and the National Science Foundation.

\section*{Data Availability}

Data used in this paper are publicly available in the Zenodo repository (\href{https://doi.org/10.5281/zenodo.22705187}{10.5281/zenodo.22705187})



\bibliographystyle{mnras}
\bibliography{main_mnras.bbl} 

\begin{thebibliography}{}
\makeatletter
\relax
\def\mn@urlcharsother{\let\do\@makeother \do\$\do\&\do\#\do\^\do\_\do\%\do\~}
\def\mn@doi{\begingroup\mn@urlcharsother \@ifnextchar [ {\mn@doi@} {\mn@doi@[]}}
\def\mn@doi@[#1]#2{\def\@tempa{#1}\ifx\@tempa\@empty \href {http://dx.doi.org/#2} {doi:#2}\else \href {http://dx.doi.org/#2} {#1}\fi \endgroup}
\def\mn@eprint#1#2{\mn@eprint@#1:#2::\@nil}
\def\mn@eprint@arXiv#1{\href {http://arxiv.org/abs/#1} {{\tt arXiv:#1}}}
\def\mn@eprint@dblp#1{\href {http://dblp.uni-trier.de/rec/bibtex/#1.xml} {dblp:#1}}
\def\mn@eprint@#1:#2:#3:#4\@nil{\def\@tempa {#1}\def\@tempb {#2}\def\@tempc {#3}\ifx \@tempc \@empty \let \@tempc \@tempb \let \@tempb \@tempa \fi \ifx \@tempb \@empty \def\@tempb {arXiv}\fi \@ifundefined {mn@eprint@\@tempb}{\@tempb:\@tempc}{\expandafter \expandafter \csname mn@eprint@\@tempb\endcsname \expandafter{\@tempc}}}

\bibitem[\protect\citeauthoryear{Aghanim et~al.,}{Aghanim et~al.}{2020}]{Aghanim2020}
Aghanim N.,  et~al., 2020, \mn@doi [Astronomy and Astrophysics] {10.1051/0004-6361/201833910}, 641, A6

\bibitem[\protect\citeauthoryear{Aihara et~al.,}{Aihara et~al.}{2011}]{Aihara2011}
Aihara H.,  et~al., 2011, \mn@doi [Astrophysical Journal, Supplement Series] {10.1088/0067-0049/193/2/29}, 193, 29

\bibitem[\protect\citeauthoryear{Arnaud, Arnaud, A., Arnaud  \& A.}{Arnaud et~al.}{1996}]{Arnaud1996}
Arnaud K.~A.,  Arnaud A. K.,  Arnaud  A. K.,  1996, ASPC, 101, 17

\bibitem[\protect\citeauthoryear{Aryan et~al.,}{Aryan et~al.}{2025}]{Aryan2025}
Aryan A.,  et~al., 2025, \mn@doi [The Astrophysical Journal Supplement Series] {10.3847/1538-4365/adfc69}, 281, 20

\bibitem[\protect\citeauthoryear{Ascenzi et~al.,}{Ascenzi et~al.}{2019}]{Ascenzi2019}
Ascenzi S.,  et~al., 2019, \mn@doi [Monthly Notices of the Royal Astronomical Society] {10.1093/mnras/stz891}, 486, 672

\bibitem[\protect\citeauthoryear{Bauer et~al.,}{Bauer et~al.}{2017}]{Bauer2017}
Bauer F.~E.,  et~al., 2017, \mn@doi [Monthly Notices of the Royal Astronomical Society] {10.1093/mnras/stx417}, 467, 4841

\bibitem[\protect\citeauthoryear{Becerra et~al.,}{Becerra et~al.}{2026}]{Becerra2026}
Becerra R.~L.,  et~al., 2026, \mn@doi [Astronomy {\&} Astrophysics] {10.1051/0004-6361/202557612}, 705, A233

\bibitem[\protect\citeauthoryear{Berger et~al.,}{Berger et~al.}{2002}]{Berger2002}
Berger E.,  et~al., 2002, \mn@doi [The Astrophysical Journal] {10.1086/344262}, 581, 981

\bibitem[\protect\citeauthoryear{Bloom, Kulkarni  \& Djorgovski}{Bloom et~al.}{2002}]{Bloom2002}
Bloom J.~S.,  Kulkarni S.~R.,   Djorgovski S.~G.,  2002, \mn@doi [The Astronomical Journal] {10.1086/338893}, 123, 1111

\bibitem[\protect\citeauthoryear{Bradley et~al.,}{Bradley et~al.}{2025}]{Bradley2025}
Bradley L.,  et~al., 2025, astropy/photutils: 2.2.0, \mn@doi{10.5281/zenodo.14889440}, \url {https://doi.org/10.5281/zenodo.14889440}

\bibitem[\protect\citeauthoryear{Brightman, Jaimes, Stern  \& Grefenstette}{Brightman et~al.}{2026}]{Brightman2026}
Brightman M.,  Jaimes J.~C.,  Stern D.,   Grefenstette B.,  2026, \mn@doi [The Astrophysical Journal, Volume 999, Issue 1, id.75, 11 pp.] {10.3847/1538-4357/ae3f31}, 999, 75

\bibitem[\protect\citeauthoryear{Brown et~al.,}{Brown et~al.}{2013}]{Brown2013}
Brown T.~M.,  et~al., 2013, \mn@doi [Publications of the Astronomical Society of the Pacific] {10.1086/673168}, 125, 1031

\bibitem[\protect\citeauthoryear{{Burrows} et~al.,}{{Burrows} et~al.}{2005}]{Burrows2005}
{Burrows} D.~N.,  et~al., 2005, \mn@doi [\ssr] {10.1007/s11214-005-5097-2}, \href {http://ads.nao.ac.jp/abs/2005SSRv..120..165B} {120, 165}

\bibitem[\protect\citeauthoryear{Campana et~al.,}{Campana et~al.}{2012}]{Campana2012}
Campana S.,  et~al., 2012, \mn@doi [Monthly Notices of the Royal Astronomical Society] {10.1111/j.1365-2966.2012.20428.x}, 421, 1697

\bibitem[\protect\citeauthoryear{Campana et~al.,}{Campana et~al.}{2014}]{Campana2014}
Campana S.,  et~al., 2014, \mn@doi [Monthly Notices of the Royal Astronomical Society] {10.1093/mnras/stu831}, 441, 3634

\bibitem[\protect\citeauthoryear{{Cardelli}, {Clayton}  \& {Mathis}}{{Cardelli} et~al.}{1989}]{Cardelli1989}
{Cardelli} J.~A.,  {Clayton} G.~C.,   {Mathis} J.~S.,  1989, \mn@doi [\apj] {10.1086/167900}, \href {http://ads.nao.ac.jp/abs/1989ApJ...345..245C} {345, 245}

\bibitem[\protect\citeauthoryear{{Cash}}{{Cash}}{1979}]{Cash1979}
{Cash} W.,  1979, \mn@doi [\apj] {10.1086/156922}, \href {https://ui.adsabs.harvard.edu/abs/1979ApJ...228..939C} {228, 939}

\bibitem[\protect\citeauthoryear{Chrimes et~al.,}{Chrimes et~al.}{2019}]{Chrimes2019}
Chrimes A.~A.,  et~al., 2019, \mn@doi [Monthly Notices of the Royal Astronomical Society] {10.1093/mnras/stz1039}, 486, 3105

\bibitem[\protect\citeauthoryear{Covino et~al.,}{Covino et~al.}{2013}]{Covino2013}
Covino S.,  et~al., 2013, \mn@doi [Monthly Notices of the Royal Astronomical Society] {10.1093/mnras/stt540}, 432, 1231

\bibitem[\protect\citeauthoryear{Dey et~al.,}{Dey et~al.}{2019}]{Dey2019}
Dey A.,  et~al., 2019, \mn@doi [The Astronomical Journal] {10.3847/1538-3881/ab089d}, 157, 168

\bibitem[\protect\citeauthoryear{Dhillon et~al.,}{Dhillon et~al.}{2018}]{Dhillon2018}
Dhillon V.,  et~al., 2018, in Ground-based and Airborne Instrumentation for Astronomy VII. p. 107020L (\mn@eprint {arXiv} {1807.00557}), \mn@doi{10.1117/12.2312041}, \url {http://adsabs.harvard.edu/abs/2018SPIE10702E..0LD}

\bibitem[\protect\citeauthoryear{Dhillon et~al.,}{Dhillon et~al.}{2021}]{Dhillon2021}
Dhillon V.~S.,  et~al., 2021, \mn@doi [Monthly Notices of the Royal Astronomical Society] {10.1093/mnras/stab2130}, 507, 350

\bibitem[\protect\citeauthoryear{Eikenberry et~al.,}{Eikenberry et~al.}{2004}]{Eikenberry2004}
Eikenberry S.~S.,  et~al., 2004, \mn@doi [Ground-based Instrumentation for Astronomy] {10.1117/12.549796}, 5492, 1196

\bibitem[\protect\citeauthoryear{{Evans} et~al.,}{{Evans} et~al.}{2009}]{Evans2009}
{Evans} P.~A.,  et~al., 2009, \mn@doi [\mnras] {10.1111/j.1365-2966.2009.14913.x}, \href {http://ads.nao.ac.jp/abs/2009MNRAS.397.1177E} {397, 1177}

\bibitem[\protect\citeauthoryear{Eyles-Ferris et~al.,}{Eyles-Ferris et~al.}{2025}]{Eyles-Ferris2025}
Eyles-Ferris R. A.~J.,  et~al., 2025, \mn@doi [The Astrophysical Journal Letters] {10.3847/2041-8213/ade1d9}, 17, 20

\bibitem[\protect\citeauthoryear{Foight, G{\"{u}}ver, {\"{O}}zel  \& Slane}{Foight et~al.}{2016}]{Foight2016}
Foight D.~R.,  G{\"{u}}ver T.,  {\"{O}}zel F.,   Slane P.~O.,  2016, \mn@doi [The Astrophysical Journal] {10.3847/0004-637x/826/1/66}, 826, 66

\bibitem[\protect\citeauthoryear{Fong et~al.,}{Fong et~al.}{2022}]{Fong2022}
Fong W.-f.,  et~al., 2022, \mn@doi [The Astrophysical Journal] {10.3847/1538-4357/ac91d0}, 940, 56

\bibitem[\protect\citeauthoryear{Fraija, Galv{\'{a}}, Kamenetskaia  \& Dainotti}{Fraija et~al.}{2026}]{Fraija2026}
Fraija N.,  Galv{\'{a}} A.,  Kamenetskaia B.~B.,   Dainotti M.~G.,  2026, MNRAS, 000, 1

\bibitem[\protect\citeauthoryear{{Gehrels} et~al.,}{{Gehrels} et~al.}{2004}]{Gehrels2004}
{Gehrels} N.,  et~al., 2004, \mn@doi [\apj] {10.1086/422091}, \href {https://ui.adsabs.harvard.edu/abs/2004ApJ...611.1005G} {611, 1005}

\bibitem[\protect\citeauthoryear{Goad et~al.,}{Goad et~al.}{2007}]{Goad2007}
Goad M.~R.,  et~al., 2007, \mn@doi [Astronomy and Astrophysics] {10.1051/0004-6361:20078436}, 476, 1401

\bibitem[\protect\citeauthoryear{Gordon \& Arnaud}{Gordon \& Arnaud}{2021}]{Gordon2021}
Gordon C.,  Arnaud K.,  2021, ASCL, p. ascl:2101.014

\bibitem[\protect\citeauthoryear{Greiner et~al.,}{Greiner et~al.}{2011}]{Greiner2011}
Greiner J.,  et~al., 2011, \mn@doi [Astronomy and Astrophysics] {10.1051/0004-6361/201015458}, 526, 30

\bibitem[\protect\citeauthoryear{Guevel, Hosseinzadeh, Bostroem  \& Burke}{Guevel et~al.}{2021}]{Guevel2021}
Guevel D.,  Hosseinzadeh G.,  Bostroem A.,   Burke C.~J.,  2021, dguevel/PyZOGY: v0.0.2, \mn@doi{doi:10.5281/zenodo.4570234}, \url {https://doi.org/doi:10.5281/zenodo.4570234}

\bibitem[\protect\citeauthoryear{Guo, Zeng, Wei, Zhou, Jin, Wu  \& Wei}{Guo et~al.}{2025}]{Guo2025}
Guo Y.,  Zeng H.,  Wei J.,  Zhou H.,  Jin Z.,  Wu X.,   Wei D.,  2025, \mn@doi [The Astrophysical Journal Letters] {10.3847/2041-8213/ae2745}, 995, L53

\bibitem[\protect\citeauthoryear{G{\"{u}}ver \& {\"{O}}zel}{G{\"{u}}ver \& {\"{O}}zel}{2009}]{Guver2009}
G{\"{u}}ver T.,  {\"{O}}zel F.,  2009, \mn@doi [Monthly Notices of the Royal Astronomical Society] {10.1111/j.1365-2966.2009.15598.x}, 400, 2050

\bibitem[\protect\citeauthoryear{Heasarc}{Heasarc}{2014}]{Heasarc2014}
Heasarc N. H. E. A. S. A. R.~C.,  2014, Ascl, p. ascl:1408.004

\bibitem[\protect\citeauthoryear{Hook, J{\o}rgensen, Allington‐Smith, Davies, Metcalfe, Murowinski  \& Crampton}{Hook et~al.}{2004}]{Hook2004}
Hook I.,  J{\o}rgensen I.,  Allington‐Smith J.,  Davies R.,  Metcalfe N.,  Murowinski R.,   Crampton D.,  2004, \mn@doi [Publications of the Astronomical Society of the Pacific] {10.1086/383624}, 116, 425

\bibitem[\protect\citeauthoryear{Izzo \& Malesani}{Izzo \& Malesani}{2024}]{gcnIzzo2024}
Izzo L.,  Malesani D.~B.,  2024, GCN, 38053, 1

\bibitem[\protect\citeauthoryear{Jakobsson, Hjorth, Fynbo, Watson, Pedersen, Bj{\"{o}}rnsson  \& Gorosabel}{Jakobsson et~al.}{2004}]{Jakobsson2004}
Jakobsson P.,  Hjorth J.,  Fynbo J. P.~U.,  Watson D.,  Pedersen K.,  Bj{\"{o}}rnsson G.,   Gorosabel J.,  2004, \mn@doi [The Astrophysical Journal] {10.1086/427089}, 617, L21

\bibitem[\protect\citeauthoryear{Jiang et~al.,}{Jiang et~al.}{2025}]{Jiang2025}
Jiang S.-Q.,  et~al., 2025, \mn@doi [The Astrophysical Journal Letters] {10.3847/2041-8213/addebf}, 988, L34

\bibitem[\protect\citeauthoryear{Johnson, Leja, Conroy  \& Speagle}{Johnson et~al.}{2021}]{Johnson2021}
Johnson B.~D.,  Leja J.,  Conroy C.,   Speagle J.~S.,  2021, \mn@doi [The Astrophysical Journal Supplement Series] {10.3847/1538-4365/abef67}, 254, 22

\bibitem[\protect\citeauthoryear{Jonker et~al.,}{Jonker et~al.}{2013}]{Jonker2013}
Jonker P.~G.,  et~al., 2013, \mn@doi [Astrophysical Journal] {10.1088/0004-637X/779/1/14}, 779, 14

\bibitem[\protect\citeauthoryear{Jonker et~al.,}{Jonker et~al.}{2025a}]{Jonker2026}
Jonker P.~G.,  et~al., 2025a, \mn@doi [Monthly Notices of the Royal Astronomical Society] {10.1093/mnras/staf2021}, 545, 1

\bibitem[\protect\citeauthoryear{Jonker, Martin-Carrillo, Malesani, Quirola-V{\'{a}}squez, Levan  \& Bauer}{Jonker et~al.}{2025b}]{Jonker2025a}
Jonker P.,  Martin-Carrillo A.,  Malesani D.~B.,  Quirola-V{\'{a}}squez J.~A.,  Levan A.~J.,   Bauer F.~E.,  2025b, GCN, 38063

\bibitem[\protect\citeauthoryear{Karambelkar, De, Mo, Ibrahim  \& Schiminovich}{Karambelkar et~al.}{2026}]{gcnKarambelkar2026}
Karambelkar V.,  De K.,  Mo G.,  Ibrahim S.,   Schiminovich D.,  2026, GCN, 44296, 1

\bibitem[\protect\citeauthoryear{Kr{\"{u}}hler et~al.,}{Kr{\"{u}}hler et~al.}{2011}]{Kruhler2011}
Kr{\"{u}}hler T.,  et~al., 2011, \mn@doi [Astronomy and Astrophysics] {10.1051/0004-6361/201117428}, 534, A108

\bibitem[\protect\citeauthoryear{Labrie et~al.,}{Labrie et~al.}{2023}]{Labrie2023}
Labrie K.,  et~al., 2023, \mn@doi [Research Notes of the AAS] {10.3847/2515-5172/ad0044}, 7, 214

\bibitem[\protect\citeauthoryear{Li et~al.,}{Li et~al.}{2024}]{gcnLi2024}
Li W.~X.,  et~al., 2024, GCN, 38054, 1

\bibitem[\protect\citeauthoryear{Li, Sun, Qian  \& Li}{Li et~al.}{2026a}]{Li2026b}
Li Q.-M.,  Sun Q.-B.,  Qian S.-B.,   Li F.-X.,  2026a, \mn@doi [The Astrophysical Journal Letters] {10.3847/2041-8213/ae2012}, 997, L15

\bibitem[\protect\citeauthoryear{Li, Wu, Xin, Qiu, Ma  \& Yao}{Li et~al.}{2026b}]{gcnLi2026a}
Li H.~L.,  Wu C.,  Xin L.~P.,  Qiu Y.,  Ma Y.,   Yao Z.,  2026b, GCN, 44244, 1

\bibitem[\protect\citeauthoryear{Li, Huang, Cao, Jiang  \& Zhang}{Li et~al.}{2026c}]{gcnLi2026b}
Li Z.~X.,  Huang G.~L.,  Cao J.~Y.,  Jiang S.,   Zhang W.,  2026c, GCN, 44251, 1

\bibitem[\protect\citeauthoryear{Lyman et~al.,}{Lyman et~al.}{2017}]{Lyman2017}
Lyman J.~D.,  et~al., 2017, \mn@doi [Monthly Notices of the Royal Astronomical Society] {10.1093/mnras/stx220}, 467, 1795

\bibitem[\protect\citeauthoryear{Magnier et~al.,}{Magnier et~al.}{2020}]{Magnier2020}
Magnier E.~A.,  et~al., 2020, \mn@doi [The Astrophysical Journal Supplement Series] {10.3847/1538-4365/abb82a}, 251, 6

\bibitem[\protect\citeauthoryear{Melandri et~al.,}{Melandri et~al.}{2012}]{Melandri2012}
Melandri A.,  et~al., 2012, \mn@doi [Monthly Notices of the Royal Astronomical Society] {10.1111/j.1365-2966.2011.20398.x}, 421, 1265

\bibitem[\protect\citeauthoryear{Metzger}{Metzger}{2020}]{Metzger2020}
Metzger B.~D.,  2020, \mn@doi [Living Reviews in Relativity] {10.1007/s41114-019-0024-0}, 23, 1

\bibitem[\protect\citeauthoryear{Morrison \& McCammon}{Morrison \& McCammon}{1983}]{Morrison1983}
Morrison R.,  McCammon D.,  1983, \mn@doi [The Astrophysical Journal] {10.1086/161102}, 270, 119

\bibitem[\protect\citeauthoryear{O'Connor et~al.,}{O'Connor et~al.}{2022}]{OConnor2022}
O'Connor B.,  et~al., 2022, \mn@doi [Monthly Notices of the Royal Astronomical Society] {10.1093/mnras/stac1982}, 515, 4890

\bibitem[\protect\citeauthoryear{O'Connor et~al.,}{O'Connor et~al.}{2025}]{OConnor2025}
O'Connor B.,  et~al., 2025, \mn@doi [The Astrophysical Journal Letters] {10.3847/2041-8213/ae146b}, 993, L37

\bibitem[\protect\citeauthoryear{Perley et~al.,}{Perley et~al.}{2009}]{Perley2009}
Perley D.~A.,  et~al., 2009, \mn@doi [Astronomical Journal] {10.1088/0004-6256/138/6/1690}, 138, 1690

\bibitem[\protect\citeauthoryear{Perley, Niino, Tanvir, Vergani  \& Fynbo}{Perley et~al.}{2016a}]{Perley2016a}
Perley D.~A.,  Niino Y.,  Tanvir N.~R.,  Vergani S.~D.,   Fynbo J.~P.,  2016a, \mn@doi [Space Science Reviews] {10.1007/s11214-016-0237-4}, 202, 111

\bibitem[\protect\citeauthoryear{Perley et~al.,}{Perley et~al.}{2016b}]{Perley2016}
Perley D.~A.,  et~al., 2016b, \mn@doi [The Astrophysical Journal] {10.3847/0004-637x/817/1/7}, 817, 7

\bibitem[\protect\citeauthoryear{Price-Whelan et~al.,}{Price-Whelan et~al.}{2022}]{Price-Whelan2022}
Price-Whelan A.~M.,  et~al., 2022, \mn@doi [The Astrophysical Journal] {10.3847/1538-4357/ac7c74}, 935, 167

\bibitem[\protect\citeauthoryear{Quirola-V{\'{a}}squez et~al.,}{Quirola-V{\'{a}}squez et~al.}{2022}]{Quirola-Vasquez2022}
Quirola-V{\'{a}}squez J.,  et~al., 2022, \mn@doi [Astronomy and Astrophysics] {10.1051/0004-6361/202243047}, 663, A168

\bibitem[\protect\citeauthoryear{Quirola-V{\'{a}}squez et~al.,}{Quirola-V{\'{a}}squez et~al.}{2023}]{Quirola-Vasquez2023a}
Quirola-V{\'{a}}squez J.,  et~al., 2023, \mn@doi [Astronomy and Astrophysics] {10.1051/0004-6361/202345912}, 675, A44

\bibitem[\protect\citeauthoryear{Quirola-V{\'{a}}squez et~al.,}{Quirola-V{\'{a}}squez et~al.}{2026}]{gcnQuirolaVasquez2026}
Quirola-V{\'{a}}squez J.,  et~al., 2026, GCN, 44264, 1

\bibitem[\protect\citeauthoryear{Ror, Pandey, Oates, Gupta, Aryan, Castro-Tirado  \& Kumar}{Ror et~al.}{2025}]{Ror2025}
Ror A.~K.,  Pandey S.~B.,  Oates S.~R.,  Gupta R.,  Aryan A.,  Castro-Tirado A.~J.,   Kumar S.,  2025, \mn@doi [Monthly Notices of the Royal Astronomical Society] {10.1093/mnras/staf1514}, 543, 2404

\bibitem[\protect\citeauthoryear{Sari, Piran  \& Narayan}{Sari et~al.}{1998}]{Sari1998}
Sari R.,  Piran T.,   Narayan R.,  1998, \mn@doi [The Astrophysical Journal] {10.1086/311269}, 497, L17

\bibitem[\protect\citeauthoryear{Simpson, Smirnova, Labrie, Hirst, Berke, Rawlings, Turner  \& Vacca}{Simpson et~al.}{2023}]{Simpson2023}
Simpson C.,  Smirnova O.,  Labrie K.,  Hirst P.,  Berke D.,  Rawlings M.,  Turner J.,   Vacca W.,  2023, {DRAGONS}, \mn@doi{https://doi.org/10.5281/zenodo.4025470}, \url {https://zenodo.org/records/17412281}

\bibitem[\protect\citeauthoryear{Skrutskie et~al.,}{Skrutskie et~al.}{2006}]{Skrutskie2006}
Skrutskie M.~F.,  et~al., 2006, \mn@doi [The Astronomical Journal] {10.1086/498708}, 131, 1163

\bibitem[\protect\citeauthoryear{Soderberg et~al.,}{Soderberg et~al.}{2008}]{Soderberg2008}
Soderberg A.~M.,  et~al., 2008, \mn@doi [Nature] {10.1038/nature06997}, 453, 469

\bibitem[\protect\citeauthoryear{Srinivasaragavan et~al.,}{Srinivasaragavan et~al.}{2025}]{Srinivasaragavan2025}
Srinivasaragavan G.~P.,  et~al., 2025, \mn@doi [arXiv] {10.48550/ARXIV.2512.10239}, 34, arXiv:2512.10239

\bibitem[\protect\citeauthoryear{Swain et~al.,}{Swain et~al.}{2026}]{Swain2026}
Swain V.,  et~al., 2026, \mn@doi [The Astrophysical Journal Letters] {10.3847/2041-8213/ae2a20}, 996, L38

\bibitem[\protect\citeauthoryear{Urata et~al.,}{Urata et~al.}{2007}]{Urata2007}
Urata Y.,  et~al., 2007, \mn@doi [Publications of the Astronomical Society of Japan] {10.1093/pasj/59.4.L29}, 59, L29

\bibitem[\protect\citeauthoryear{Valenti et~al.,}{Valenti et~al.}{2017}]{Valenti2017}
Valenti S.,  et~al., 2017, \mn@doi [The Astrophysical Journal Letters] {10.3847/2041-8213/aa8edf}, 848, L24

\bibitem[\protect\citeauthoryear{Volnova et~al.,}{Volnova et~al.}{2014}]{Volnova2014}
Volnova A.~A.,  et~al., 2014, \mn@doi [Monthly Notices of the Royal Astronomical Society] {10.1093/mnras/stu999}, 442, 2586

\bibitem[\protect\citeauthoryear{Watson}{Watson}{2011}]{Watson2011}
Watson D.,  2011, \mn@doi [Astronomy and Astrophysics] {10.1051/0004-6361/201117120}, 533, A16

\bibitem[\protect\citeauthoryear{Yuan, Zhang, Chen  \& Ling}{Yuan et~al.}{2022}]{Yuan2022}
Yuan W.,  Zhang C.,  Chen Y.,   Ling Z.,  2022, \mn@doi [Handbook of X-ray and Gamma-ray Astrophysics] {10.1007/978-981-16-4544-0_151-1}, pp 1--30

\bibitem[\protect\citeauthoryear{Yuan et~al.,}{Yuan et~al.}{2025}]{Yuan2025}
Yuan W.,  et~al., 2025, \mn@doi [Science China: Physics, Mechanics and Astronomy] {10.1007/s11433-024-2600-3}, 68, 239501

\bibitem[\protect\citeauthoryear{Zackay, Ofek  \& Gal-Yam}{Zackay et~al.}{2016}]{Zackay2016}
Zackay B.,  Ofek E.~O.,   Gal-Yam A.,  2016, \mn@doi [The Astrophysical Journal] {10.3847/0004-637x/830/1/27}, 830, 27

\bibitem[\protect\citeauthoryear{Zand, Guidorzi, Heise, Amati, Kuulkers, Frontera, Gianfagna  \& Piro}{Zand et~al.}{2025}]{Zand2025}
Zand J. J. M. i.~t.,  Guidorzi C.,  Heise J.,  Amati L.,  Kuulkers E.,  Frontera F.,  Gianfagna G.,   Piro L.,  2025, arXiv

\bibitem[\protect\citeauthoryear{Zhang, Sun, Yin, Yang, Zhang  \& Wu}{Zhang et~al.}{2024}]{Zhang2024b}
Zhang B.,  Sun H.,  Yin Y.-H.~I.,  Yang J.,  Zhang B.,   Wu X.,  2024, Astronomer's Telegram, 16473, 1

\bibitem[\protect\citeauthoryear{Zhao, Liang, Liu, Zhang  \& Collaborations}{Zhao et~al.}{2024a}]{Zhao2024}
Zhao T.,  Liang Y.~F.,  Liu H.~Y.,  Zhang W.,   Collaborations t. E.~P.,  2024a, GCN, 38051, 1

\bibitem[\protect\citeauthoryear{Zhao, Liang, Liu, Zhang  \& Collaborations}{Zhao et~al.}{2024b}]{Zhao2024b}
Zhao T.,  Liang Y.~F.,  Liu H.~Y.,  Zhang W.,   Collaborations t. E.~P.,  2024b, Gcn, 38058, 1

\bibitem[\protect\citeauthoryear{Zheng, Deng  \& Wang}{Zheng et~al.}{2009}]{Zheng2009}
Zheng W.-K.,  Deng J.-S.,   Wang J.,  2009, \mn@doi [Research in Astronomy and Astrophysics] {10.1088/1674-4527/9/10/003}, 9, 1103

\bibitem[\protect\citeauthoryear{van Dalen et~al.,}{van Dalen et~al.}{2025}]{vanDalen2025}
van Dalen J. N.~D.,  et~al., 2025, \mn@doi [The Astrophysical Journal Letters] {10.3847/2041-8213/adbc7e}, 982, L47

\bibitem[\protect\citeauthoryear{van Hoof et~al.,}{van Hoof et~al.}{2026}]{vanHoof2026c}
van Hoof A. P.~C.,  et~al., 2026, MNRAS, 000, 1

\bibitem[\protect\citeauthoryear{van~der Horst, Kouveliotou, Gehrels, Rol, Wijers, Cannizzo, Racusin  \& Burrows}{van~der Horst et~al.}{2009}]{vanderHorst2009}
van~der Horst A.~J.,  Kouveliotou C.,  Gehrels N.,  Rol E.,  Wijers R.~A.,  Cannizzo J.~K.,  Racusin J.,   Burrows D.~N.,  2009, \mn@doi [Astrophysical Journal] {10.1088/0004-637X/699/2/1087}, 699, 1087

\makeatother
\end{thebibliography}





\bsp	
\label{lastpage}
\end{document}